\documentclass[useAMS,usenatbib]{mn2e}
\usepackage{amssymb}
\usepackage{epsfig}

\def\mnras{MNRAS}  % Monthly Notices of the RAS
\def\apj{ApJ}      % The Astrophysical Journal
\def\apjl{ApJL}    % The Astrophysical Journal Letters
\def\apjs{ApJS}    % The Astrophysical Journal Supplement
\def\aap{A\&A}     % Astronomy and Astrophysics
\def\aaps{A\&AS}   % Astronomy and Astrophysics Supplement
\def\araa{ARA\&A}  % Annual Reviews of Astronomy and Astrophysics
\def\nat{Nature}   % Nature
\def\owls{{\small OWLS} }
\def\pasp{PASP}
\def\pasj{PASJ}

\title[AGN feedback in galaxy groups]{The case for AGN feedback in galaxy groups}

\author[I. G. McCarthy et al.]{I. G. McCarthy$^{1,2,3}$\thanks{E-mail:
mccarthy@ast.cam.ac.uk (IGM)}, J. Schaye$^4$, T. J. Ponman$^5$, R. G. Bower$^6$, C. M. Booth$^4$, \newauthor
C. Dalla Vecchia$^{4,7}$, R. A. Crain$^8$, V. Springel$^9$, T. Theuns$^{6,10}$, R. P. C. Wiersma$^{4}$
\\
\\
$^{1}$Kavli Institute for Cosmology, University of Cambridge, Madingley Road, Cambridge, CB3 OHA\\
$^{2}$Astrophysics Group, Cavendish Laboratory, JJ Thomson Avenue, Cambridge, CB3 0HE\\
$^{3}$Institute of Astronomy, University of Cambridge, Madingley Road, Cambridge, CB3 0HA\\ 
$^{4}$Leiden Observatory, Leiden University, P. O. Box 9513, 2300 RA Leiden, the Netherlands\\
$^{5}$Astrophysics and Space Research Group, School of Physics and Astronomy, University of Birmingham, Edgbaston, Birmingham, B15 2TT\\ 
$^{6}$Institute of Computational Cosmology, Department of Physics, University of Durham, Science Laboratories, South Road, Durham DH1 3LE\\
$^7$Max Planck Institute for Extraterrestrial Physics, Giessenbachstrabe 1, 85748 Garching, Germany\\
$^8$Centre for Astrophysics \& Supercomputing, Swinburne University of Technology, Hawthorn, Victoria 3122, Australia\\
$^9$Max-Planck-Institute for Astrophysics, Karl-Schwarzschild-Str. 1, 85741, Garching, Germany\\
$^{10}$Department of Physics, University of Antwerp, Campus Groenenborger, Groenenborgerlaan 171, B-2020 Antwerp, Belgium
}

\begin{document}

\date{Accepted XXXX. Received XXXX; in original form XXXX}

\pagerange{\pageref{firstpage}--\pageref{lastpage}} \pubyear{2009}

\maketitle

\label{firstpage}

\begin{abstract}

The relatively recent insight that energy input from supermassive black holes (BHs) can have a substantial effect on the star formation rates (SFRs) of galaxies motivates us to examine the effects of BH feedback on the scale of galaxy groups.  At present, groups contain most of the galaxies and a significant fraction of the overall baryon content of the universe and, along with massive clusters, they represent the only systems for which it is possible to measure both the stellar and gaseous baryonic components directly.  To explore the effects of BH feedback on groups, we analyse two high resolution cosmological hydrodynamic simulations from the OverWhelmingly Large Simulations (OWLS) project.  While both include galactic winds driven by supernovae, only one of the models includes feedback from accreting BHs.  We compare the properties of the simulated galaxy groups to a wide range of observational data, including the entropy and temperature profiles of the intragroup medium, hot gas mass fractions, the luminosity$-$temperature and mass$-$temperature scaling relations, the K-band luminosity of the group and its central brightest galaxy (CBG), star formation rates and ages of the CBG, and gas- and stellar-phase metallicities.  Both runs yield entropy distributions similar to the data, while the run without AGN feedback yields highly peaked temperature profiles, in discord with the observations.  Energy input from supermassive BHs significantly reduces the gas mass fractions of galaxy groups with masses less than a few times $10^{14}$ M$_\odot$, yielding a gas mass fraction and X-ray luminosity scaling with system temperature that is in excellent agreement with the data, although the detailed scatter in the $L-T$ relation is not quite correct.  The run without AGN feedback suffers from the well known overcooling problem --- the resulting stellar mass fractions are several times larger than observed and present-day cooling flows operate uninhibitedly.  By contrast, the run that includes BH feedback yields stellar mass fractions, SFRs, and stellar age distributions in excellent agreement with current estimates, thus resolving the long standing `cooling crisis' of simulations on the scale of groups.  Both runs yield very similar gas-phase metal abundance profiles that match X-ray measurements, but they predict very different stellar metallicities.  Based on the above, galaxy groups provide a compelling case that feedback from supermassive BHs is a crucial ingredient in the formation of massive galaxies.

\end{abstract}


\section{Introduction}

Galaxy formation is one of the key unsolved problems in modern astrophysics.  Traditionally, studies of galaxy formation have focused on the observable stellar properties of galaxies.  However, a large fraction of the baryonic mass of massive galaxies is believed to be in a hot diffuse form and many of the processes that regulate the formation of stars do so by influencing the properties of this hot gas.  Thus, a complete view of galaxy formation necessarily incorporates both the stars and the hot gas and an understanding of the processes by which these phases interact.

At present, galaxy groups and clusters represent the only systems in the universe for which it is possible to measure the properties of both the stellar and gaseous baryonic components out to a significant fraction of the halo viral radius.  This is possible because the gaseous component is compressed and heated to very high temperatures ($T \sim 10^{6-8}$ K) where this material (the intragroup or intracluster media) becomes visible at X-ray wavelengths.  While clusters have higher X-ray surface brightnesses than groups, and the derived physical properties of the gas are therefore presumably more robust, clusters represent very rare and extreme environments.  Since the efficiencies of many of the environmental processes that are thought to alter galaxy formation/evolution (e.g., ram pressure stripping, strangulation, tidal stripping, galaxy harassment) increase rapidly with the mass of the main halo, one expects that galaxies in clusters have been subjected to far more violent interactions than those which reside in the lower density and more typical environment of groups.  It is therefore possible that a picture of galaxy formation derived in large part from observations of clusters is misleading.

Galaxy groups, on the other hand, contain a large fraction of present day galaxies and a significant fraction of the overall universal baryon budget (e.g., Mulchaey 2000).  They are therefore more representative hosts than clusters.  In addition, the effects of feedback from supernovae (SNe) and supermassive black holes (BHs) on the hot gas, which is thought to be crucial in shaping the properties of galaxies, are expected to be much more obvious (and therefore perhaps easier to quantify) on the scale of groups, where the energy input associated with these sources is comparable to the binding energies of these systems.  And while what could be said about the detailed properties of the hot gas was severely limited in the past (with data from the previous generation of X-ray satellites), new high-quality data from the {\it Chandra} and {\it XMM-Newton} X-ray telescopes has yielded a much more detailed picture of the thermodynamic state of the gas in groups.  For these reasons, galaxy groups probably represent the best objects for studying gas and stars simultaneously.

Much progress has also been made in the theoretical modeling of the formation and evolution of groups and clusters.  For example, the past two decades have seen the development of sophisticated cosmological hydrodynamic simulation codes capable of simulating the properties of the hot gas in groups and clusters {\it ab initio} with millions to hundreds of millions of particles/cells (e.g., Evrard 1990; Thomas \& Couchman 1992; Navarro et al.\ 1995; Bryan \& Norman 1998; Kravtsov et al.\ 2000; Borgani et al.\ 2001; Springel et al.\ 2001; Kay et al.\ 2002).  Broadly speaking, these simulations have had varying degrees of success in reproducing the thermodynamic properties of the hot gas but, in general, those simulations that included the effects of radiative cooling appear to have formed far too many stars relative to what is observed in local groups and clusters (see Balogh et al.\ 2001 for further discussion of this `cooling crisis').  This may be because, until very recently, such simulations did not include feedback from supermassive BHs.  In recent years, a growing number of authors have argued, based on the results semi-analytic models of galaxy formation as well as hydrodynamic simulations (usually idealised, as opposed to cosmological), that feedback from supermassive BHs plays a crucial role in regulating the star formation rates of massive galaxies (e.g., Springel et al.\ 2005, hereafter SDH05; Bower et al.\ 2006, 2008; Croton et al.\ 2006; Hopkins et al.\ 2006; Somerville et al.\ 2008) and suppressing the onset of catastrophic cooling of the hot gas in groups and clusters (e.g,. Churazov et al.\ 2001; Brighenti \& Mathews 2003; Dalla Vecchia et al.\ 2004; Ruszkowski et al.\ 2004; Omma et al.\ 2004).

The insight that AGN feedback is important in the formation and evolution of massive systems motivated SDH05 (see also Thacker et al.\ 2006) to develop a novel scheme for following the growth of supermassive BHs and feedback from AGN self-consistently in cosmological smoothed particle hydrodynamics (SPH) simulations.  Using the model of SDH05 as a backbone, Sijacki \& Springel (2006), Sijacki et al.\ (2007), Puchwein et al.\ (2008), Bhattacharya et al.\ (2009), and Fabjan et al.\ (2009) examined the effects of AGN feedback\footnote{Note that Sijacki et al.\ (2007), Puchwein et al.\ (2008), and Fabjan et al.\ (2009) modified the original implementation of SDH05 to include both `quasar' and `radio' feedback modes.  The former is characterised by isotropic thermal heating of the gas when accretion rates are high, while the latter, which is imposed when accretion rates are low, is characterised by thermal heating of bi-polar spherical regions offset from the galaxy hosting the BH, which are meant to mimic `bubbles' observed in many nearby clusters.} on groups and clusters.  These authors indeed find that including self-consistent AGN feedback yields reduced stellar mass fractions and star formation rates while reproducing a number of key hot gas observable properties as well.  For example, Puchwein et al.\ (2008) found that including energy input from supermassive BHs enables their simulations to reproduce the luminosity-temperature and gas mass fraction-temperature relations of groups and clusters.  Fabjan et al.\ (2009) found similarly good matches to these relations, while demonstrating also that the simulations yield metallicity and temperature profiles in reasonable agreement with observations of galaxy groups (although the predicted stellar fractions still appear to be somewhat too high and the entropy-temperature scaling is shallower than observed).

In the present study, we examine the effects of AGN feedback on galaxy groups using a novel set of simulations.  In particular, we extract two runs from the OverWhelmingly Large Simulations ({\small OWLS}, for short; Schaye et al.\ 2009) and attempt to ascertain whether or not feedback from supermassive BHs is a necessary ingredient for explaining the properties of observed galaxy groups.  The simulations we analyse have very different implementations for radiative cooling, star formation, chemodynamics, and feedback from supernova and AGN from those of SDH05 and the studies mentioned above which are based upon that model.  This means a comparison of the findings can be used to check the robustness of the conclusions to differences in the parameterisations of various subgrid physics as well as inclusion of new physics.  In addition, because the \owls simulations are performed in large periodic boxes (as opposed to being re-simulations using `zoomed initial conditions') they yield much larger samples of groups for analysis than used previously for simulations with AGN feedback, albeit at lower mass resolution (e.g., each \owls $100$ ${\rm h}^{-1}$ Mpc box run yields $\approx 200$ groups with masses greater than $10^{13}$ M$_\odot$, whereas the studies based upon re-simulations typically only analyse a handful of systems).  This is potentially important, as observed groups and clusters show a large degree of system-to-system scatter at fixed mass in the properties of the hot gas (e.g., Fabian et al.\ 1994; Markevitch 1998; McCarthy et al.\ 2004).  We compare the \owls simulations to a much wider range of X-ray and optical observations of galaxy groups than previously considered.  This allows for a more thorough assessment of the successes/failures of the models.  Finally, the large suite of \owls runs also allows us to explore the impact of additional sub-grid physics, not just for impact of AGN feedback.  For example, the effects of changing the supernova feedback prescription, switching off metal-line, and changing the stellar initial mass function.  These variations will be explored in a forthcoming paper.

The present paper is organised as follows.  In Section 2 we provide a brief description of the \owls runs examined in this study and how we select and analyse the galaxy groups that form in these simulations.  In Section 3 we make detailed like-with-like comparisons with a wide range of observational data of both the hot gas and stellar baryonic components.  In Section 4 we summarise and discuss our findings.  We adopt a flat $\Lambda$CDM cosmology with $h=0.73$, $\Omega_b=0.0418$, $\Omega_m=0.238$, and $\sigma_8=0.74$, which are the maximum-likelihood values from fitting to the 3-year {\it WMAP} CMB temperature anisotropy data (Spergel et al.\ 2007).  When comparing observed and simulated metal abundances, we consistently adopt the solar abundances of Asplund et al.\ (2005).

Finally, we note this is the first in a series of papers investigating galaxy groups in the \owls suite of simulations.  In Paper II we examine in detail the thermal history of the intragroup medium and present a simple physical model for the excess entropy seen at intermediate/large radii in groups.  In Paper III we explore the effect of varying the parameters of the key subgrid physics models (e.g., stellar IMF, chemodynamics, supernova and BH feedback) on the properties of galaxy groups.  In Paper IV we investigate the build up of metals in the intragroup medium and the origin of the large scale gradients.

\section{Galaxy groups in \owls}

An in-depth presentation of the \owls project can be found in Schaye et al.\ 
(2009). For the present study, we shall 
therefore give only a brief description of the simulations and our analysis methods. 
In Paper III in our series on galaxy groups, we will provide a detailed description of 
our analysis methods, including how we find and select the simulated galaxy groups and 
then produce and analyse mock observations of them.  

\subsection{Simulation characteristics}

The \owls project has been undertaken with the goal of exploring galaxy formation and 
its sensitivity to both resolvable and `sub-grid' physics in fully self-consistent 
cosmological hydrodynamic simulations.  The approach is, starting from identical initial 
conditions, to vary the key parameters of the physics modules (e.g., such as the 
adopted stellar initial mass function (IMF), the rate at which black holes accrete mass, the 
wind velocity and mass-loading of supernova driven winds, switching on/off metal-dependent 
radiative cooling, switching on/off mass/energy transfer from SNIa and AGB stars, etc.) to 
see what effects they have on the resulting population of galaxies, or in this case galaxy 
groups.  Comparisons between the various runs should allow one to isolate which
physical processes are most important for, for example, establishing the shape of the stellar mass 
function or the entropy of the intragroup medium.  

In Papers II and III in our series on galaxy groups, we will present detailed comparisons 
between the various \owls runs and focus on 
the physics of the hot gas in galaxy groups.  In the present study, however, we select 
just a subset of two \owls runs; the reference run (hereafter {\it REF}) and the default 
AGN feedback run (hereafter {\it AGN}).  These two runs are identical in every way, 
except that the latter incorporates a prescription for BH growth and AGN 
feedback, as described in Booth \& Schaye (2009) (hereafter BS09), whereas the former does not.  The goal 
of 
the present study, therefore, is to ascertain whether including feedback from AGN is 
required to explain the observable properties of galaxy groups.  

The \owls runs are initialised from 
identical initial conditions; i.e., from a $\Lambda$CDM cosmological density field (with a 
power spectrum computed using CMBFAST) in a periodic box of $100 h^{-1}$ Mpc on a 
side at $z=127$.  In generating this density field, the various relevant cosmological 
parameters ($\Omega_b$, $\Omega_m$, $\Omega_\Lambda$, $h$, $\sigma_8$, and $n_s$) were set 
equal to their maximum-likelihood values found from the analysis of the 3-year {\it WMAP} 
CMB data (Spergel et al.\ 2007).  The simulations were evolved to $z=0$ using the 
TreePM-SPH code {\small GADGET-3} (last described in Springel 2005) and we extract from the final snapshot ($z=0$) all galaxy 
groups for which M$_{200} \ge 10^{13}$ M$_\odot$ (where M$_{200}$ is the total mass within 
a radius that encloses a mean density of 200 times the present-day critical density of 
the universe).  The simulations use $2\times512^3$ particles, yielding a mass resolution for gas and dark matter particles $m_{\rm gas} 
\approx 8.65 \times 10^7 h^{-1}$ M$_\odot$ and  m$_{\rm dm} \approx 4.06 \times 10^8 h^{-1}$ 
M$_\odot$, respectively. Thus, a $\sim 10^{14} h^{-1}$ M$_\odot$ galaxy group is 
resolved with $\sim 10^5$ gas and dark matter particles.  As we demonstrate in the Appendix, this resolution is sufficient to robustly predict the properties of galaxy groups in our simulations.

The {\small GADGET-3} code has been substantially modified to incorporate new baryonic 
physics.
Radiative cooling rates for the gas are computed on an element-by-element basis by 
interpolating within pre-computed tables (generated with CLOUDY) that contain 
cooling rates as a function of density, temperature, and redshift calculated in the 
presence of the cosmic microwave background and photoionisation from a Haardt \& Madau 
(2001) ionising UV/X-Ray background (see Wiersma et al.\ 2009a).  Star formation is 
tracked in the simulations following the prescription of Schaye \& Dalla Vecchia (2008).   
Gas with densities exceeding the critical density for the onset of the thermo-gravitational 
instability is expected to be multiphase and to form stars (Schaye 2004). Because the
simulations lack both the physics and the resolution to model the cold interstellar gas 
phase, an effective equation of state (EOS) is imposed with pressure $P \propto \rho^{4/3}$  
for densities $n_H > n_*$ where $n_* = 0.1$ cm$^{-3}$.  As described in Schaye \& Dalla 
Vecchia (2008), gas on the effective EOS is allowed to form stars at a pressure-dependent
rate that reproduces the observed Kennicutt-Schmidt law (Kennicutt 1998) by construction.  
The timed release of individual elements (``metals'') by both massive (Type II SNe and 
stellar winds) and intermediate mass stars (Type Ia SNe and asymptotic giant branch stars) 
is included following the prescription of Wiersma et al.\ (2009b).  A set of 11
individual elements are followed in these simulations (H, He, C, Ca, N, O, Ne, Mg, S, Si, 
Fe), which represent all the important species for computing radiative cooling rates.

In the {\it REF} run, feedback from supernovae is incorporated using the kinetic wind model 
of Dalla Vecchia \& Schaye (2008) with the wind velocity, $v_w$, set to $600$ km/s and the 
mass-loading parameter (i.e., the ratio of the mass of gas given a velocity kick to that 
turned into newly formed star particles), $\eta$, set to $2$.  This corresponds to using 
approximately 40\% of the total energy available from supernovae for a Chabrier (2003) IMF, which is assumed by default.  This choice of parameters results in a good match to the 
peak of the star formation rate history of the universe (Schaye et al.\ 2009), but otherwise is 
somewhat arbitrary.  However, the {\it REF} run is typical, 
in terms of the galaxy group properties it produces, of all of the \owls runs that 
implement supernovae feedback only with efficiencies $\le 1$ (i.e., varying $\eta$ and 
$v_w$ does not greatly affect the resulting group properties for constant $\eta v_w^2$).  This will be demonstrated in 
a forthcoming paper.

The {\it AGN} run also includes kinetic supernovae feedback and uses the same choices for $v_w$ 
and $\eta$.  The prescription for the growth of BHs and feedback from AGN that we use, which is a substantially modified version of that introduced by SDH05, is described in detail in BS09.
A basic description is as follows.  Black holes grow by accretion of surrounding 
gas and by mergers with other black holes.  In the simulations analysed in the present study, the Bondi radius is generally not resolved.  This lack of resolution will result in an underestimate of the gas 
density near the BH, and therefore an underestimate of the BH accretion rate.  The simulations particularly underestimate the Bondi rate if the gas is multiphase, since we impose an effective equation of state for gas with $n_H > n_*$.  To compensate for this effect, the calculated Bondi-Hoyle-Lyttleton rate is multiplied by a coefficient $\alpha$, as was done originally by SDH05.  In the model of SDH05 (and the many studies that have used this model), $\alpha$ is fixed typically at a value of $\approx 100-300$, whereas BS09 instead have $\alpha$ vary as a power law that depends on the local gas density.  The power law index, $\beta$, is fixed to 2 for the model we examine in the present study and the relation is normalised such that $\alpha = 1$ when the local gas density is equal to the density threshold for star formation\footnote{As discussed in BS09, any simulation that resolves the Jeans scales automatically resolved the Bondi radius of any BH whose mass exceeds the particle mass.  Thus, normalising the power law between $\alpha$ and $\rho$ in this way ensures that $\alpha$ goes to unity when the Bondi radius is resolved.}.  This approach has the same number of free parameters (one) as the implementation of SDH05, but has the advantage that the accretion rate does not exceed the Bondi rate when the Bondi radius is resolved in the simulation.  In any circumstance, the gas accretion rate may not exceed the Eddington rate, and most of the BH mass is assembled with accretion rates at or close to the Eddington rate.

A certain fraction of the rest mass energy of the accreted gas, $\epsilon$, is used to 
heat a number of randomly selected neighbouring gas particles, $n_{\rm heat}$, by raising their temperatures by an amount $\Delta T_{\rm heat}$.  (Note that the BHs store feedback energy until it is sufficient to heat $n_{\rm heat}$ by $\Delta T_{\rm heat}$.)  This efficiency, which is actually the product of two separate 
efficiencies, namely 
the radiative efficiency of the BH accretion disk and the fraction of the emitted energy that 
is assumed to couple to the gas, is a free parameter of the model.  BS09 find that a value 
of $\epsilon=0.015$ yields a good match to the $z=0$ relations between BH mass and stellar mass and velocity dispersion and the $z=0$ cosmic BH 
density.  
We therefore fix $\epsilon$ at this value.

Aside from the efficiency, the {\it AGN} run is characterised by the following 
3 main parameters.  The index in the power law scaling of the Bondi accretion 
coefficient, $\alpha$, is set to $\beta=2$, $n_{\rm heat}=1$, and, $\Delta T_{\rm heat}=10^8$ K.  In a forthcoming paper, we show the  
properties of galaxy groups simulated with AGN feedback are relatively 
insensitive to changes in $\beta$ and $n_{\rm heat}$, 
but that the results {\it are} sensitive to the temperature increase, 
particularly if it is comparable to, or smaller than, the virial temperature of the 
galaxy group, which for our sample is $\sim 10^7$ K.  The reason for this is obvious, if 
the gas is not heated to a higher temperature than the surrounding ambient medium, then its 
entropy and cooling time will remain low.  This will rapidly lead to a pile-up of 
low entropy material in the central regions of groups and clusters, violating 
observational constraints.  We have found that a temperature increase of $\Delta T_{\rm 
heat}\ga10^8$ K resolves this problem on the group scale.

It should be noted that, as individual gas particles are selected randomly for heating, 
the mode of feedback from BHs developed by BS09 will not necessarily produce large sets of cavities (or 
`bubbles') of the kind frequently observed in massive nearby cool core clusters. (Thus, the model of BS09 does not distinguish between `quasar' and `radio' modes, unlike the model developed by Sijacki et al.\ 2007). We are 
currently developing models in which bi-polar bubbles are inflated, similar to that of Sijacki et al.\ (2007),
to see if this results in significant changes in the ability to match observations on the group scale.  
Qualitatively, though, we would expect the results to be similar, as to be successful 
in staving off catastrophic cooling in specific directions any such model will likely 
have to heat the gas over the full $4\pi$ steradians in a time-averaged sense (which happens 
by construction in the BS09 model of BH feedback).

\subsubsection{Analysis of galaxy groups}

For each $z=0$ galaxy group with $M_{200} \ge 10^{13}$ M$_\odot$, of which there are approximately 200 in both runs, we produce simple mock observations.  For the hot gas we generate synthetic X-ray 
observations, whereas for the stars we produce surface brightness maps in a 
variety of optical/near-infrared bands (e.g. Ks, B, V).  These maps are used
to i) identify the peak of the X-ray surface brightness distribution, which we designate 
as the centre of the system when constructing 2D or 3D radial profiles (such as 
density, temperature, element abundances, etc.), and ii) for the stars, to identify the 
central brightest galaxy (the CBG) and its centre.

In the case of hot gas properties, such as mean temperature or gas density profiles, we 
emission-weight the gas properties\footnote{Physical quantities derived from real X-ray observations are necessarily emission-weighted (i.e., in an annulus the most luminous gas will contribute the most to the signal).  This is why we emission-weight all of the quantities from the simulations.  In practice, though, noticeable differences in the mass- and emission-weighted profiles only occur in the outskirts of the groups, which are typically not observable anyway.  Note also that the emission-weighted temperature that we calculate is defined in precisely the same way as by Sun et al.\ (2009), to which many of our comparisons are made.  The emission-weighted temperature differs from the Mazzotta et al.\ (2004) definition of ``spectroscopic'' temperature typically by only 10\%.} so that an approximate like-with-like comparison can 
be made directly with the observations.  The X-ray luminosity of the j$^{\rm th}$ gas 
particle is computed as:
\begin{eqnarray}
L_{\rm X,j} & = & n_{e,j} n_{H,j} \Lambda_j V_j\\
           & = & \frac{X_e(Z_j)}{[X_e(Z_j)+X_i(Z_j)]^2} \biggl(\frac{\rho_j}{\mu(Z_j) m_H} \biggr)^2 V_j \nonumber \\
           & = & \frac{X_e(Z_j)}{[X_e(Z_j)+X_i(Z_j)]^2} \frac{\rho_j}{\mu(Z_j) m_H} 
                 \frac{m_{gas,j}}{\mu(Z_j) m_H} \Lambda_j \nonumber
\end{eqnarray}

\noindent where $\rho$ is the particle gas density, $m_{gas}$ is the particle mass, the volume is $V = m_{gas} / \rho$, $n_e$, $n_H$ and $n_i$ are the number densities of electrons, hydrogen, and ions, respectively, $X_e \equiv n_{\rm e}/n_{\rm H}$, $X_i \equiv n_{\rm i}/n_{\rm H}$, $Z$ is the metallicity, $\mu$ is the mean molecular weight, $m_H$ is the mass 
of a hydrogen atom, and $\Lambda$ is the cooling function in units of ergs cm$^3$ s$^{-1}$ (integrated over some 
appropriate passband, such as 0.5-2.0 keV).  We compute
$\Lambda$ by interpolating a pre-computed table generated using the Astrophysical 
Plasma Emission Code\footnote{ 
To maintain strict consistency with the implementation of radiative
cooling in the simulations, it would be more appropriate to use
cooling rates predicted by the CLOUDY software package (Ferland et al.\ 1998).  The APEC,
however, is more widely used in the analysis of X-ray data, which
is why we have adopted it here.  We have compared the rates
predicted by the APEC against those predicted by the CLOUDY and
MEKAL (another widely used plasma model) codes and, reassuringly,
we find negligible (less than a few percent) differences between
the predictions in the regime in which we are interested (i.e., hot, dense gas).}
 (APEC v1.3.1; see ${\rm http://cxc.harvard.edu/atomdb/}$).  APEC cooling rates are computed on an element-by-element 
basis and summed to yield the total cooling rate of each particle; i.e.,

\begin{equation}
\Lambda_j(T_j) = \sum_{k=1}^{N} \lambda_{j,k}(T_j)
\end{equation}

\noindent where $T$ is the gas temperature and $\lambda_{j,k}(T_j)$ is the cooling function for
element species $k$ for the {\it j}$^{\rm th}$ particle. The summation is
performed over the 11 most important elements for cooling
(H, He, C, Ca, N, O, Ne, Mg, S, Si, Fe), which are individually and self-consistently 
tracked during the simulation.  

Optical luminosities are computed by treating each star particle as a simple stellar 
population (SSP), which is reasonable as the particle mass is comparable to that of a star 
cluster.  The simulations adopt a Chabrier IMF and store the age 
and metallicity of the particles.  We use this information to compute a spectral energy 
distribution for each star particle using the GALAXEV model of Bruzual \& Charlot (2003).  
The luminosity is obtained by integrating the product of the SED with an appropriate 
transmission filter function (e.g., B, V, or $K_s$ filters).

\section{Results}

Below we make detailed comparisons of the two \owls runs with a wide variety of observational data.
In Section 3.1 we examine the entropy and temperature distributions of the hot gas.  The gas mass fractions of groups are presented in Section 3.2.  In Section 3.3 we analyse the `classical' X-ray mass-temperature and luminosity-temperature scaling relations.  In Section 3.4 we examine the overall star formation efficiency of galaxy groups, the mean ages of the CBGs, and the fraction of CBGs that are currently forming stars.  Finally, in Section 3.5 we analyse the gas- and stellar-phase metallicities of groups. 

\subsection{Entropy and temperature distributions}

\subsubsection{Entropy}

We begin by examining the entropy distribution of the gas in the {\it REF} and {\it AGN} \owls 
runs and how it compares with that of observed galaxy groups.  As is common in 
X-ray astronomy, we define the `entropy', $S$, as $k_B T/n_e^{2/3}$ and use units of keV cm$^2$.  The observational entropy $S$ is related to the 
thermodynamic specific entropy $s$ via $s \propto \ln{S}$ and, like the 
thermodynamic entropy, $S$ will be conserved in any adiabatic process.  
Therefore, unlike the temperature or density, which may be raised/lowered by 
any processes that compress/expand the gas, the entropy will maintain a record of 
the thermodynamic history of the gas (see Voit et al.\ 2003; Voit 2005).

In Fig.\ 1, we show the median 3D emission-weighted entropy 
profiles for the {\it REF} and {\it AGN} runs.  This is the median profile for all galaxy 
groups with masses in the range $13.25 \le \log_{10}($M$_{500}/$M$_\odot) \le 14.25$ 
[approximately 70 groups with a median mass of $\log_{10}($M$_{500}/$M$_\odot) \approx 13.5$], where M$_{500}$ is the total mass within $r_{500}$, which is the radius that encloses a mean density of 500 times the present-day critical density of 
the universe.  
Shown for comparison are the hatched regions, which represent the 20th and 80th  
percentiles of the {\it XMM-Newton} sample of Johnson, Ponman, \& Finoguenov (2009) [approximately 
20 groups with a median mass of $\log_{10}($M$_{500}/$M$_\odot) \approx 13.5$] and the {\it 
Chandra} sample of Sun et al.\ (2009) [approximately 40 groups with a median mass of 
$\log_{10}($M$_{500}/$M$_\odot) \approx 13.8$].  Also shown (dotted line) is the power law fit 
of Voit, Kay, \& Bryan (2005) to the entropy profiles of a sample of groups and 
clusters simulated with non-radiative physics (i.e., the self-similar answer).  The radial coordinate has been normalised by $r_{500}$. Typically, $r_{500} \approx 0.65 r_{200}$. The entropy has been normalised by $S_{500}$, the `virial entropy', which is defined as:
\begin{eqnarray}
S_{500}(z=0) & \equiv & \frac{k_B T_{500}}{[n_{e,500}]^{2/3}} \\
& = & \frac{G M_{500} \mu m_H}{2 r_{500} [500 f_b \rho_{crit}(z=0)
/ (\mu_e m_H)]^{2/3}} \nonumber
\end{eqnarray}  

\noindent where $f_b = \Omega_b/\Omega_m = 0.1756$, and $\mu_e$ is the mean molecular weight per free electron.  Note that, $S_{500}$ is {\it not} the entropy at $r_{500}$, even in the case of the self-similar model.  This is because $S_{500}$ is defined in terms of the mean electron density within $r_{500}$ (as opposed to the density at $r_{500}$) {\it for a system with the universe baryon fraction}, and the virial temperature, $k_B T_{500} \equiv G M_{500} \mu m_H / 2 r_{500}$, which is not equivalent to the gas temperature at $r_{500}$ for non-isothermal mass distributions (for an NFW distribution, e.g., $T(r_{500}) \sim 0.6 T_{500}$; e.g., Loken et al.\ 2002).  $S_{500}$ therefore depends only on the halo mass, which is dominated by dark matter, and therefore is insensitive to the thermodynamic state of the gas.

\begin{figure}
\includegraphics[scale=0.55]{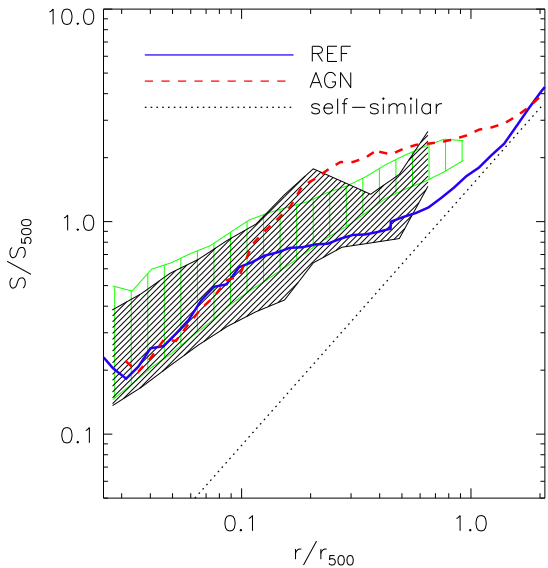}
\caption{Radial entropy distribution of the gas in the {\it REF} and {\it AGN} \owls runs.
Shown are the median emission-weighted 3D entropy profiles for the {\it REF}  
(solid
blue curve) and {\it AGN} (dashed red curve) \owls runs for all groups with $13.25 \le
\log_{10}($M$_{500}/$M$_\odot) \le 14.25$.  The hatched black and green regions
represent the {\it XMM-Newton} data of Johnson, Ponman, \& Finoguenov (2009) and the {\it Chandra} data 
of Sun et al.\ (2009), respectively.  The hatched regions enclose the 20th and 80th 
percentiles of the data.  The simulated entropy profiles are remarkably similar in the central regions, implying that BH feedback, while preventing excessive star formation in the CBG (see Section 3.4), does not significantly disturb the hot gas.}
\label{fig:entropy1}
\end{figure}

Both the observed and simulated groups exhibit a significant degree of `excess entropy' with respect to the self-similar scaling, generally confirming the results of previous studies (e.g., Finoguenov et al.\ 2002).  This 
is a tell-tale sign of the impact of non-gravitational physics (i.e., cooling and 
feedback) on the properties of the intragroup medium.  Within $r_{500}$ (at least), all of 
the gas has been significantly affected by non-gravitational physics.  This is contrary to what is seen in the most massive clusters ($T_X > 4$ keV), 
where at intermediate radii the gas reverts to the self-similar profile (McCarthy et al.\ 2008; Pratt et al.\ 2009).

At small radii ($r \la 0.1 r_{500}$), the two \owls runs yield profiles that are very 
similar and are consistent with the data, although the simulated gradient may be slightly too steep with respect to the data.  The similarity of the profiles from the two runs at small 
radii is somewhat surprising, as in the {\it AGN} run the time since the last 
feedback outburst from the central BH is typically short ($\sim 10^{7-8}$ yrs).  This 
implies that, although the heating from the AGN is sufficient to 
prevent excessive star formation in the CBG (see Section 3.4), it does not significantly 
modify the emission-weighted properties of the hot intragroup gas at radii out to $\sim 0.1 r_{500}$.  This may be consistent with the observational findings of Jetha et al.\ (2007), who found very little difference in the entropy profiles between groups with and without bright central radio sources.  The similarity of the simulated group profiles may stem from the fact that the AGN heats only small packets of gas at a time, which 
are able to float buoyantly out to larger radii without doing much work on the 
surrounding gas (see discussion in McCarthy et al.\ 2008).  

At larger radii, however, the 
two models show an interesting departure from one another (with the {\it AGN} run having 
more excess entropy) that effectively brackets the observational data.  Note, however, that since the median masses and mass distributions of the simulated and observed group samples are not identical it is difficult to assess based purely on Fig.\ 1 which model gives the best match to the data.  We address this mass dependence directly below.

\begin{figure*}
\includegraphics[scale=0.385]{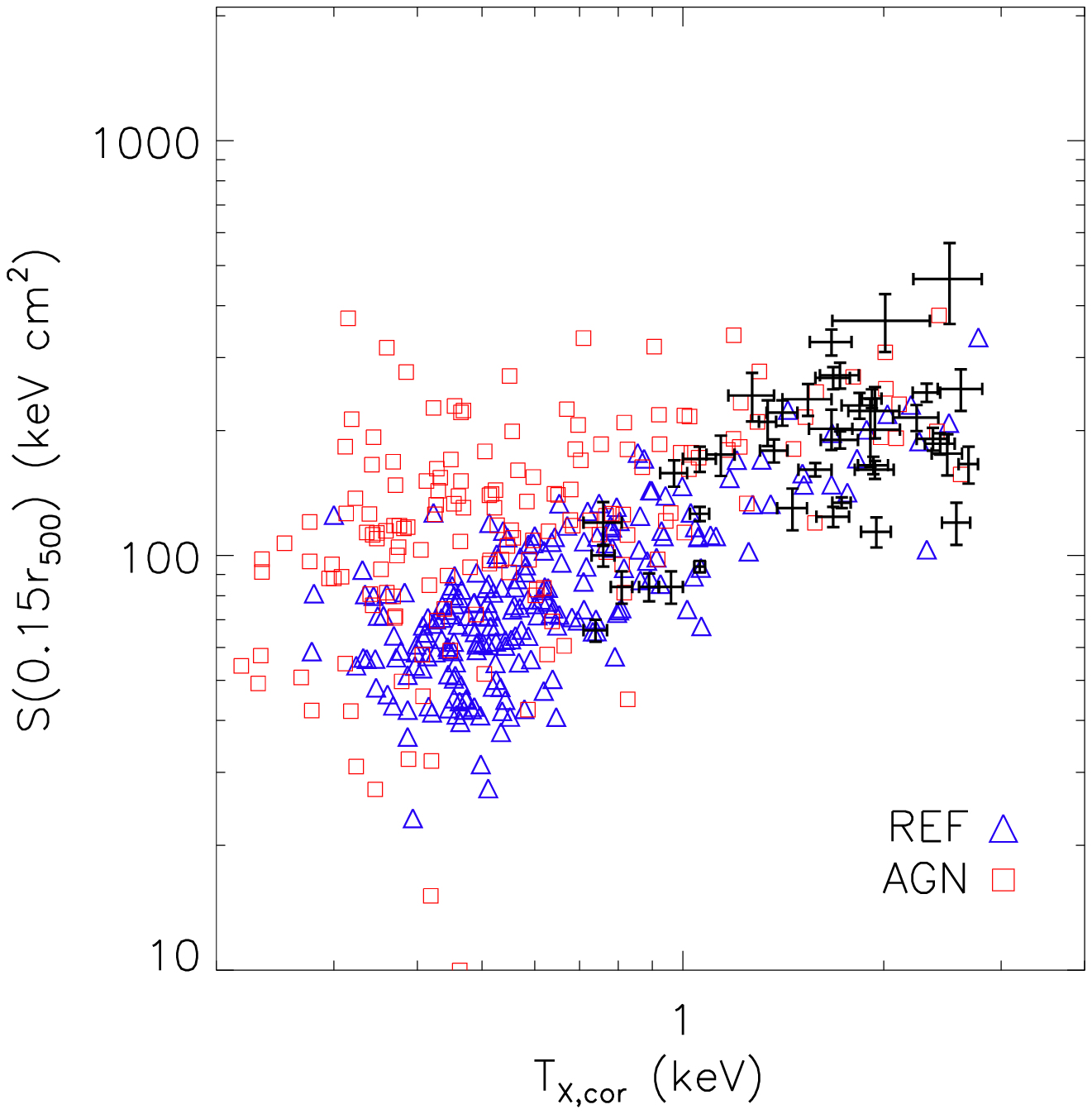}
\includegraphics[scale=0.385]{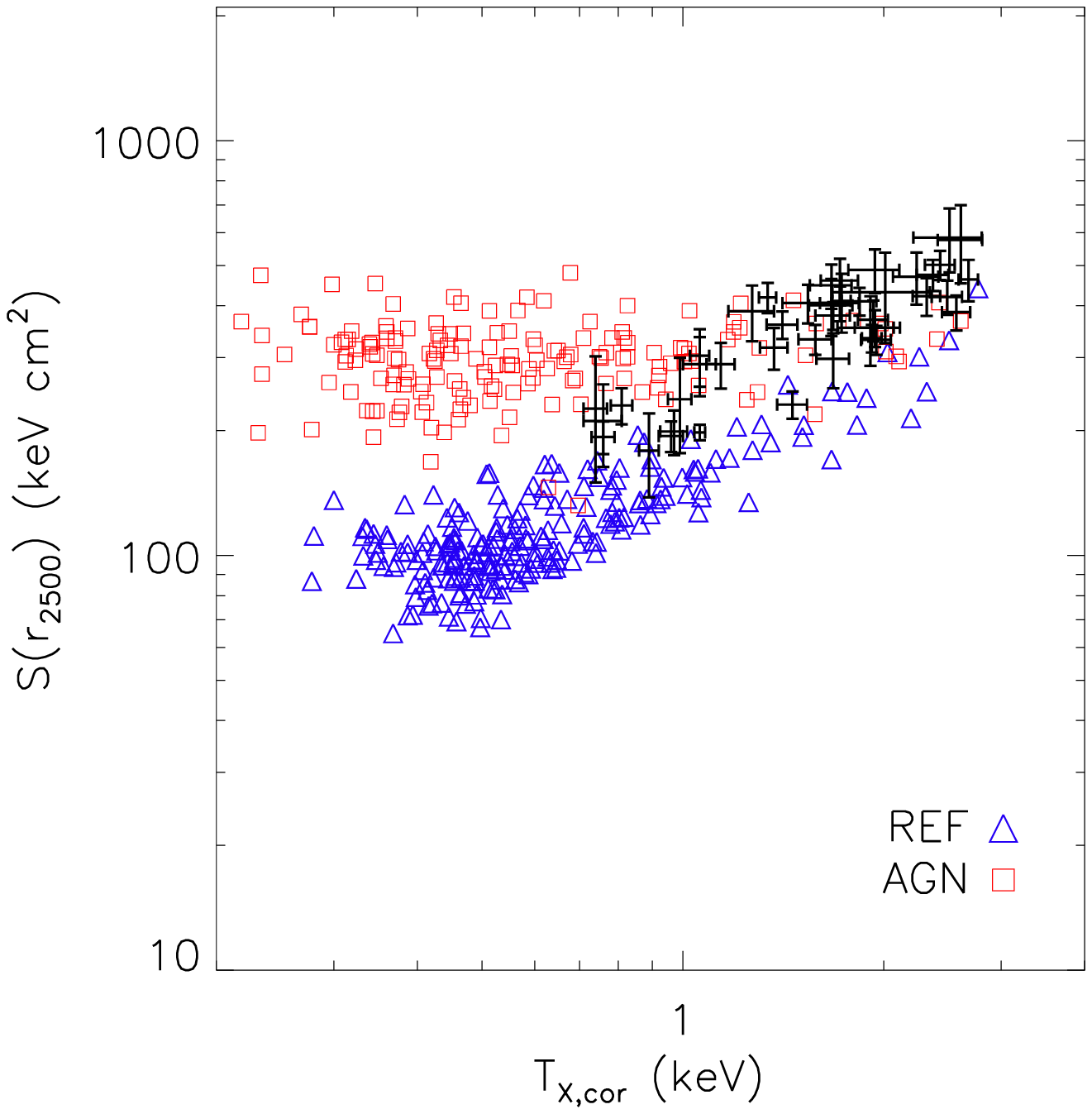}
\includegraphics[scale=0.385]{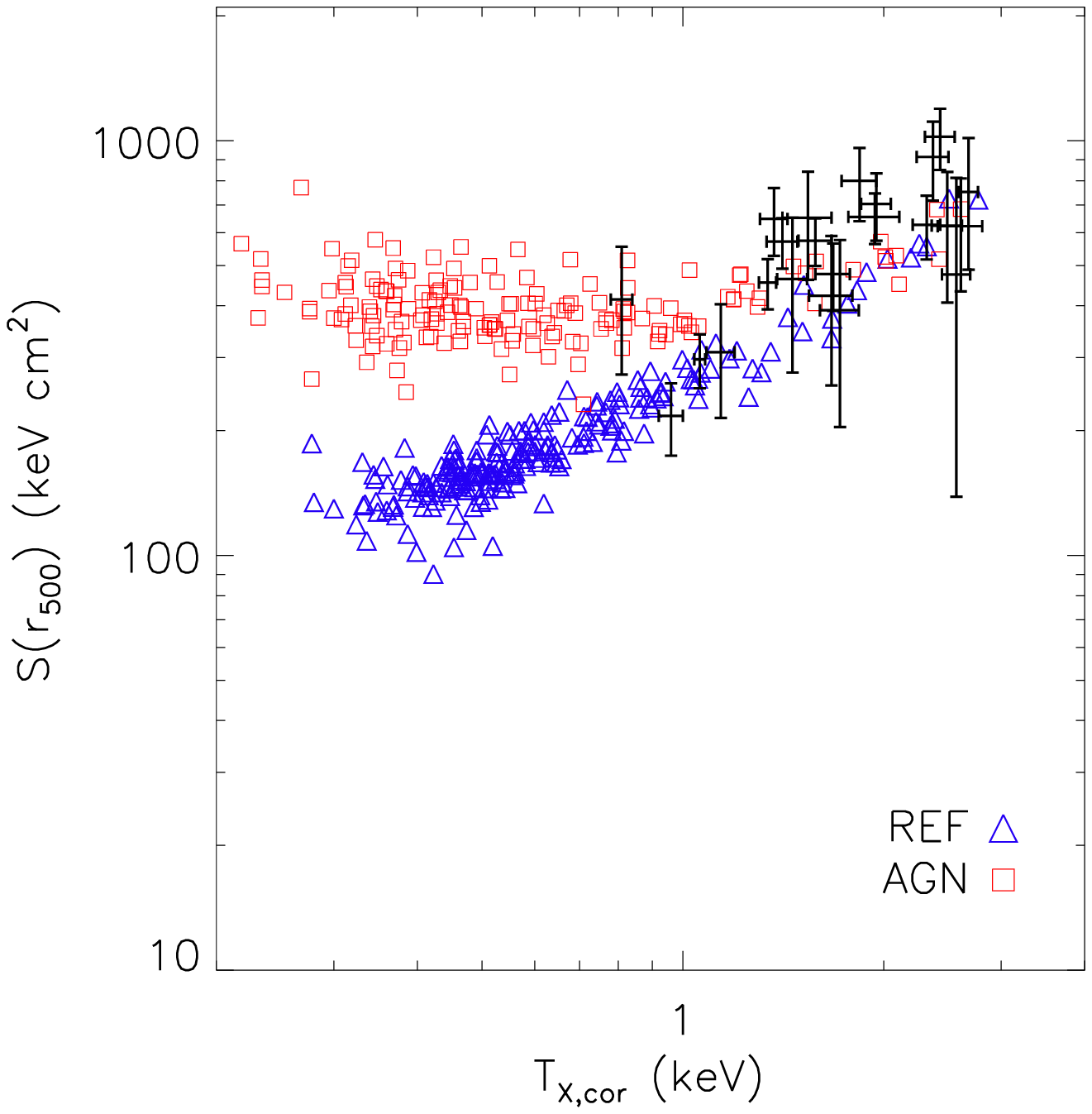}
\caption{Entropy at several different characteristic radii as a function of cool core-corrected emission-weighted system temperature.  {\it Left:} Entropy at $0.15 r_{500}$.   {\it Middle:} Entropy at $r_{2500}$.  {\it Right:} Entropy at $r_{500}$.  The error bars represent the measurements of Sun et al.\ (2009).  The differences between the two \owls runs are largest for systems with $T_X \la 1$ keV, for which there are currently few entropy measurements.  
}
\label{fig:entropy2}
\end{figure*}

While enlightening, median profiles remove information about trends with system mass and scatter at fixed mass.  One way to examine the trends with mass and scatter is to plot the entropy at several fixed characteristic radii versus system mass, or some proxy for system mass such as the mean X-ray temperature (see, e.g., Ponman et al.\ 2003).  In Fig.\ 2, we show the entropy measured at $0.15 r_{500}$, $r_{2500}$, and 
$r_{500}$ vs.\ mean X-ray temperature (note that $r_{2500} \approx 0.45 r_{500}$) for all $\approx 200$ simulated groups with $M_{200} > 10^{13}$ M$_\odot$.  We make a like-with-like comparison with the data of Sun et al.\ (2009) by 
emission-weighting the entropy of the simulated groups and computing `cool core 
corrected' emission-weighted mean temperatures (see Section 3.3 for a discussion of this `cool 
core correction'), by excising the gas within projected radii of $0.15 r_{500}$ when computing the mean temperature.

Consistent with the median profile results, we find that the two models are very similar 
at small radii (left panel) and consistent with the data.  Interestingly, there 
is a much larger degree of scatter in the entropy at small radii in 
the {\it AGN} run, which will manifest itself later as increased scatter in the 
luminosity-temperature relation (see Section 3.3) and which perhaps also gives rise to scatter in the 
present-day star formation rate of the CBG (see Section 3.4).  At larger radii, the two models 
begin to depart, with the {\it AGN} model yielding a marginally better match to the Sun et 
al.\ data, particularly at intermediate radii.  Note that this slightly better match to the data is not as apparent in the median profiles plotted in Fig.\ 1, which is a result of small differences in the median masses and mass distributions of the simulated and observed group samples that were stacked to create the median profiles.

The difference between the {\it AGN} and {\it REF} runs at intermediate/large radii is greatest for small system masses where presently there is not much observational data available.  
Extending the observed samples to lower masses would therefore be very useful for helping 
to discriminate between the models, although the intrinsically low surface brightnesses of such systems will not make such measurements easy by any means.

Based on the above, we find that the effect of AGN feedback on the 
entropy distribution of the intragroup medium is rather modest (even though, as we will 
see, AGN have a large effect on the ability of the gas to form stars), with the 
largest effects actually occurring at large radii.  However, such 
similar behaviour in the entropy is actually what is {\it required} to 
match the observations, as models with no AGN feedback (such as the {\it REF} model) 
match the observed entropy distribution relatively well.  On the basis of 
entropy alone, therefore, there is not a compelling case for the need for AGN feedback.  

Our findings are broadly consistent with the results of several previous studies (e.g., Muanwong et al.\ 2002), who found that either excessive cooling {\it or} strong feedback can raise the entropy of the ICM to a level comparable to what is observed (the former mechanism does so by selectively removing the lowest-entropy gas via star formation; see Bryan 2000; Voit \& Bryan 2001).  However, it is still somewhat of a surprise that the detailed radial entropy profiles of the {\it REF} and {\it AGN} runs are so similar.  From these results alone it is not possible to distinguish what physical mechanism has caused the loss of the low entropy material.

\begin{figure}
\includegraphics[scale=0.55]{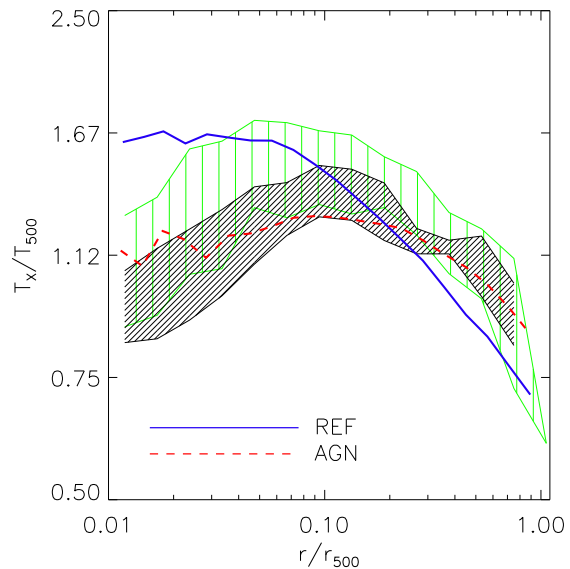}
\caption{
Median emission-weighted 2D (projected) temperature profiles.  The shaded black and green 
regions represent {\it Chandra} data of Rasmussen \& Ponman (2009) and Sun
et al.\ (2009), respectively, enclosing the 20$^{\rm th}$ and 80$^{\rm th}$ percentiles 
of their data.  The simulated profiles depart within $r \la 0.1 r_{500}$ due to the excessive accumulation of cold baryons in the centres of groups in the run with supernova feedback only.
}
\label{fig:Tgradient}
\end{figure}

\subsubsection{Temperature}

In Fig.\ 3, we show the median 2D (projected) emission-weighted temperature 
profiles for the {\it REF} and {\it AGN} runs.  These are the median profiles for all galaxy 
groups with masses in the range $13.25 \le \log_{10}($M$_{500}/$M$_\odot) \le 14.25$.  
Shown for comparison are the hatched regions, which represent the 20th and 80th  
percentiles of the {\it Chandra} sample of Rasmussen \& Ponman (2009) [15 groups with a median mass of $\log_{10}($M$_{500}/$M$_\odot) \approx 13.5$] and the {\it 
Chandra} sample of Sun et al.\ (2009) (see Section 3.1.1 for description of this sample).

\begin{figure*}
\includegraphics[scale=0.55]{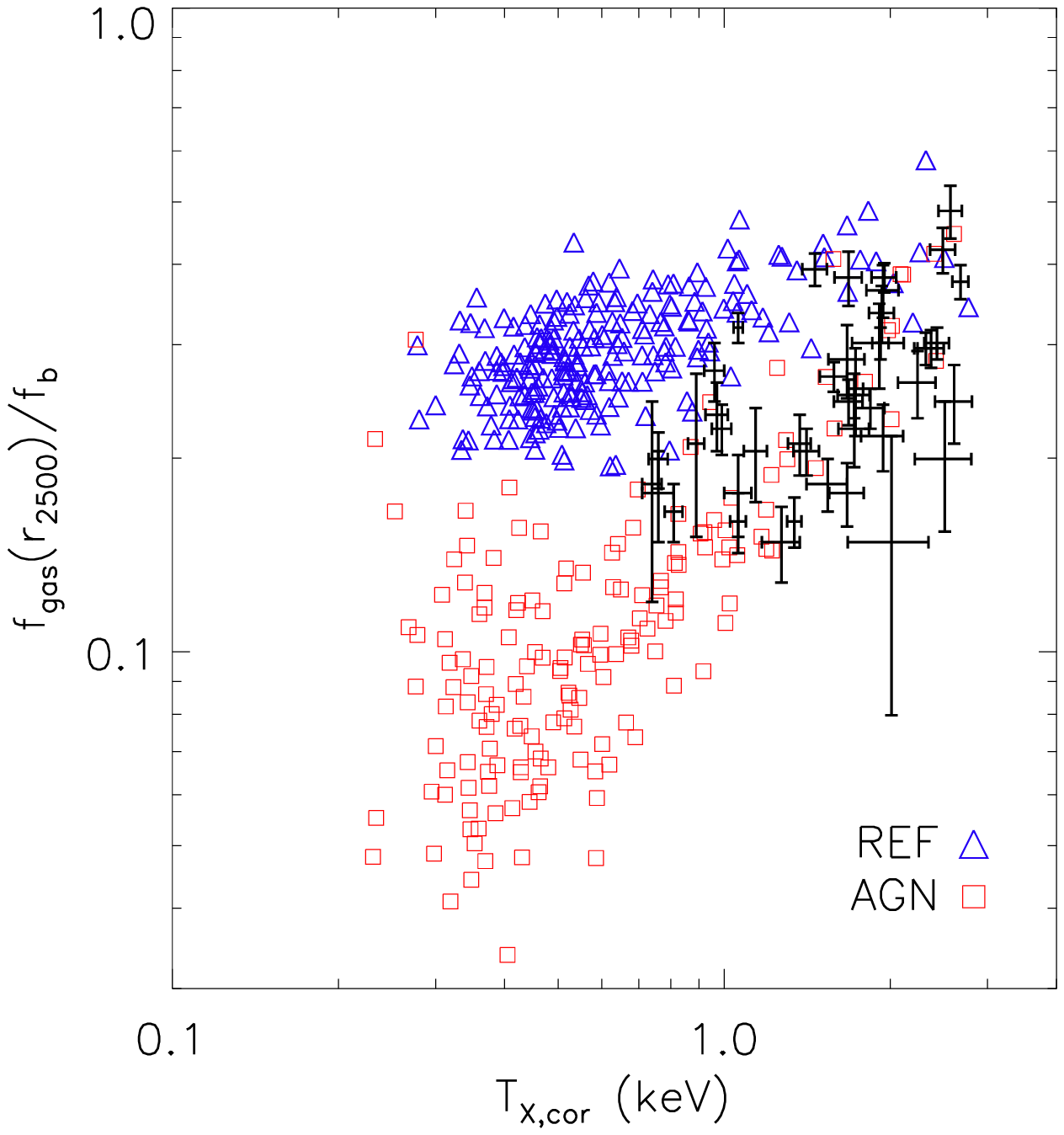}
\includegraphics[scale=0.55]{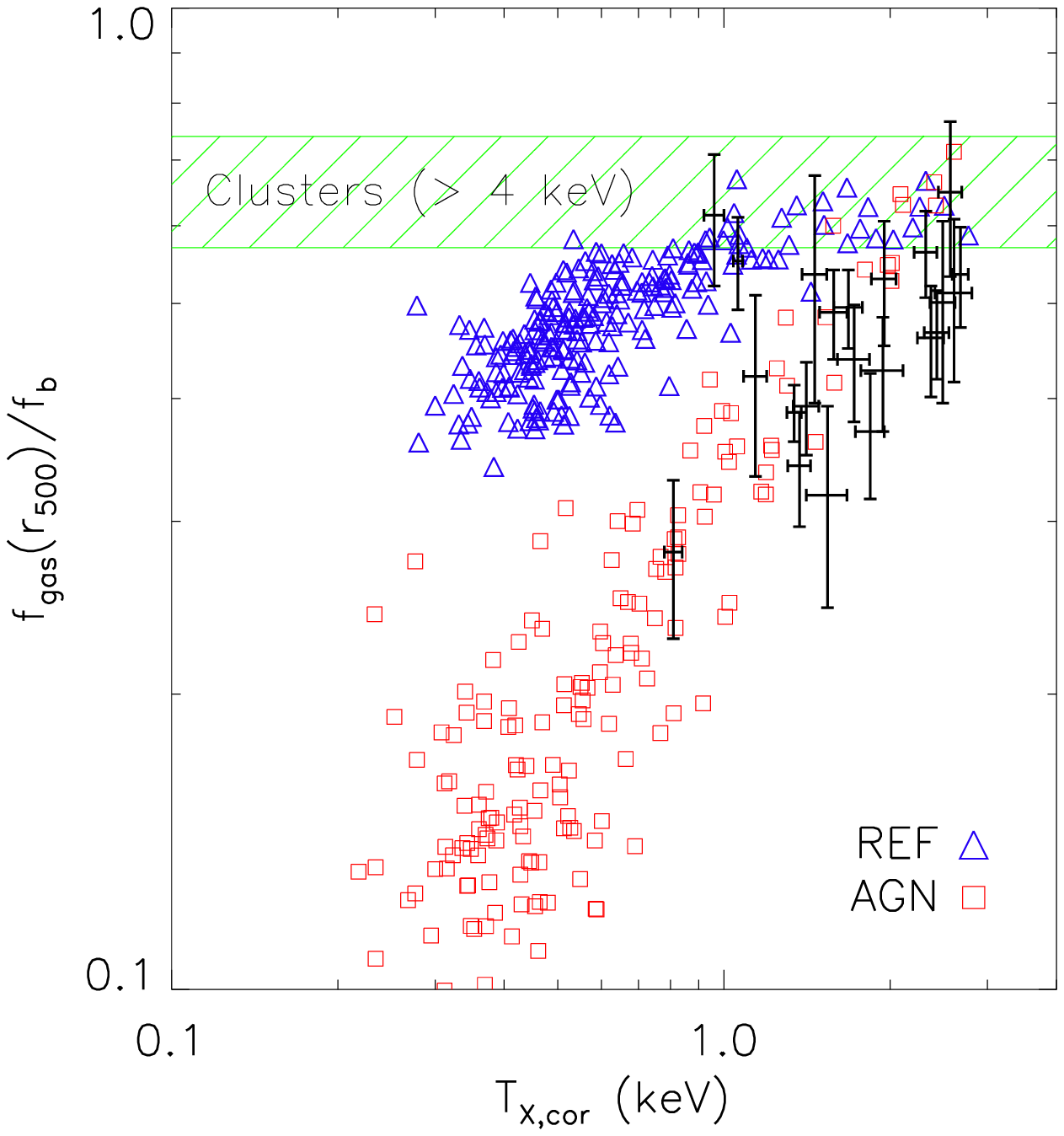}
\caption{The gas mass fraction of galaxy groups as a function of cool core-corrected   
emission-weighted system temperature.  {\it Left:} The gas mass
fraction within $r_{2500}$ [i.e., M$_{\rm gas}(r_{2500})/$M$_{2500}$] normalised by
the universal value, $f_b \equiv \Omega_b/\Omega_m = 0.18$ vs.\ cool core-corrected
emission-weighted temperature.  {\it Right:} The
analogous plot for $r_{500}$.  The hatched region represents the spread in gas mass fractions within $r_{500}$ for massive clusters (see McCarthy et al.\ 2007).
The error bars in both panels represent the {\it Chandra}-derived measurements of Sun et
al.\ (2009).  Energy input from supermassive BHs significantly reduces the gas mass fractions of groups with M$_{500} \la 10^{14}$ M$_\odot$, bringing them into agreement with the observations.
}
\label{fig:fgas}
\end{figure*}

Through hydrostatic equilibrium (HSE), the gas temperature distribution can be thought of as a manifestation of the underlying entropy profile and the depth of the gravitational potential well (see discussion in Voit 2005).  If dark matter dominates the gravitational potential well and if its distribution is relatively insensitive to what the baryons are doing, one expects that similar entropy distributions will give rise to similar temperature distributions.  Indeed, we see that at large radii ($r \ga 0.1 r_{500}$) the temperature distributions from the two runs are similar, and characterised by a mild negative gradient that is also seen in the observed groups.  The {\it AGN} run yields a slightly better fit to the data at large radii.  

Interestingly, the runs produce rather different temperature profiles within the central regions ($r \la 0.1 r_{500}$).
Since the entropy distributions at small radii are very similar (Fig.\ 1) and since the gas is very close to HSE in both runs at small radii (we have verified this), this can only signal that the temperature difference is due to a difference in the depth of the potential wells in the very central regions.  In particular, increasing the amount of mass in the central regions will give rise to additional compressional adiabatic `heating' (a rise in temperature) as gas flows into the centre.  The higher central temperatures in the {\it REF} run therefore implies that there has been a much larger degree of accumulation of (cold) baryons (likely accompanied by a contraction of the dark matter halo) at the centres of groups in this run, than in the {\it AGN} run.

The observed groups typically show evidence for a positive temperature gradient in the very central regions.  If one assumes that the dark matter distribution of the observed systems is similar to the simulated groups (e.g., Gastaldello et al.\ 2007), and given that the entropy distribution is similar to the simulated groups, the implication is that fewer cold baryons have accumulated in the central regions of real groups than predicted by the simulation that includes supernova feedback only.  This is a prediction we explicitly test in Section 3.4.  The run that includes feedback from BHs, however, yields central temperatures that are similar to observed, although the positive gradient appears to be somewhat too shallow.  Based on this better agreement, we therefore predict that this run should also yield much more sensible stellar masses within the central regions of groups.

Finally, based on the above we would argue that the presence/absence of a central positive gradient in the temperature profile is a poor proxy for the importance of radiative cooling, at least for simulations.  The entropy would be a better quantity for assessing the importance of cooling, as it is unaffected by adiabatic compression.  Which entropy threshold one adopts for characterising `cool core' and `non-cool core' systems is somewhat arbitrary, but this is unimportant as long as the same threshold is adopted for both the simulations and observations.

\subsection{Gas mass fractions}

Another important property that is sensitive to the efficiency of radiative cooling 
and feedback processes is the gas mass fraction, $f_{\rm gas}$ (i.e., the ratio of the mass of hot gas 
to total mass within some characteristic radius).  Until recently, it was only possible 
to directly measure the gas mass fractions in groups out to relatively small radii.  
However, deeper observations, particularly with {\it Chandra}, are now allowing 
measurements of the gas mass fractions of groups as far out as $r_{500}$ (e.g., 
Gastaldello et al.\ 2007; Rasmussen \& Ponman 2007; Sun et al.\ 2009).  This 
gives a much larger lever arm for discriminating between thermal histories of 
the intragroup medium and it also allows one to compare groups to clusters at the characteristic radius where clusters show evidence for convergence in $f_{\rm gas}$ (see, 
e.g., Vikhlinin et al.\ 2006; McCarthy et al.\ 2007).

In Fig.\ 4, we plot the gas mass fractions of groups in the {\it REF} and {\it AGN} \owls 
runs within two characteristic radii, $r_{2500}$ and $r_{500}$, vs.\ cool core-corrected 
emission-weighted temperature.  For comparison we also show {\it 
Chandra}-derived gas mass fractions from Sun et al.\ (2009).  The observed groups show  
very strong trends in $f_{\rm gas}$ with system temperature (mass) at both $r_{2500}$ and 
$r_{500}$, with the lowest mass systems having the lowest gas mass fractions.  
The predicted $f_{\rm gas}-T_X$ relation from the {\it AGN} run is in 
excellent agreement with the data at both $r_{2500}$ and $r_{500}$.  The {\it 
REF} (no AGN feedback) run, by contrast, shows a much milder trend in $f_{\rm gas}$ 
with system temperature and it is only for the most massive systems that simulated groups 
resemble the observed ones.  

These results appear to agree very well with those derived recently by Puchwein et al.\ (2008) and Fabjan et al.\ (2009), who also implemented BH growth and AGN feedback in self-consistent `zoomed' cosmological simulations of a number of groups and massive clusters.  This agreement comes in spite of the many differences in detailed implementations of cooling, star formation, chemodynamics, SNe winds, and AGN feedback in the \owls simulations and those of Puchwein et al.\ (2008) and Fabjan et al.\ (2009).

A simple qualitative explanation for these results is that additional energy input from 
supermassive BHs is sufficient to blow baryons out of low mass haloes or, 
alternatively, to raise the entropy of the gas to a high enough level to prevent it from 
ever being fully accreted by the system.  The effect of feedback is largest, in a 
relative sense, for low mass haloes, which have the smallest gravitational binding 
energies.  The gas mass fractions of the two \owls runs start to overlap at the highest 
system masses, for M$_{500} \ga 10^{14}$ M$_\odot$.  This therefore likely signals the mass 
scale where even supermassive BHs do not inject sufficient energy into the gas to prevent 
systems from accreting nearly\footnote{To be exact, one must 
also account for the stellar mass fraction to find the halo mass at which systems 
become `baryonically-closed'.  This occurs at approximately M$_{500} \sim 3\times10^{14}$ 
M$_\odot$ in the {\it AGN} run (and for M$_{500} \ga 
10^{13}$ M$_\odot$ in the {\it REF} run).} their full compliment of baryons.  This 
agrees remarkably well with the semi-analytic AGN model results of Bower et al.\ (2008).

\begin{figure}
\includegraphics[scale=0.55]{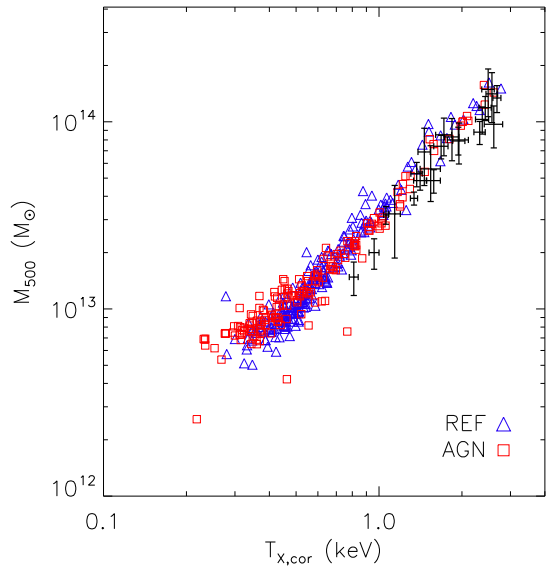}
\caption{The total mass within $r_{500}$ vs.\ cool core-corrected
emission-weighted temperature.  Error bars represent the {\it Chandra}
measurements of Sun et al.\ (2009).  The two \owls runs predict very similar relations.
 }
\label{fig:mt_relation}
\end{figure}

\subsection{Classical X-ray scaling relations}

\subsubsection{Mass-temperature relation}

Fig.\ 5 shows the predicted relations between the total mass within $r_{500}$ 
and the cool core-corrected emission-weighted mean temperature for the {\it REF} and {\it 
AGN} runs.  Also shown is the observed relationship of Sun et al.\ (2009), derived by 
applying a hydrostatic analysis to {\it Chandra} data of a sample of approximately 
40 groups.  Overall, 
both models reproduce the observed relationship relatively well, especially the slope and 
scatter.  These results are consistent with the findings of several previous studies based on 
cosmological simulations, including Borgani et al.\ (2004) and Nagai et al.\ (2007).

There is perhaps a hint of a slight $\sim10\%$ deviation in the normalisation of 
simulated and observed relations.  One possible explanation for this deviation is that the 
observed masses may be biased slightly low due to a break down in the assumption of strict HSE.  Nagai et al.\ (2007) have shown that if M$_{500}$ is 
measured by applying the assumption of HSE to cosmologically-simulated groups, they 
retrieve a M$_{500}-T_X$ relation that matches the observed relation of massive 
clusters better than in the case where they use the true masses from the simulations.  
These authors have shown that non-thermal pressure support, in the form of turbulent 
gas motions, is present at the $\sim10-20\%$ (of thermal pressure) level at radii of 
$r_{500}$ and beyond and this is what leads to the slight bias in the observed mass estimates.

\begin{figure*}
\includegraphics[scale=0.55]{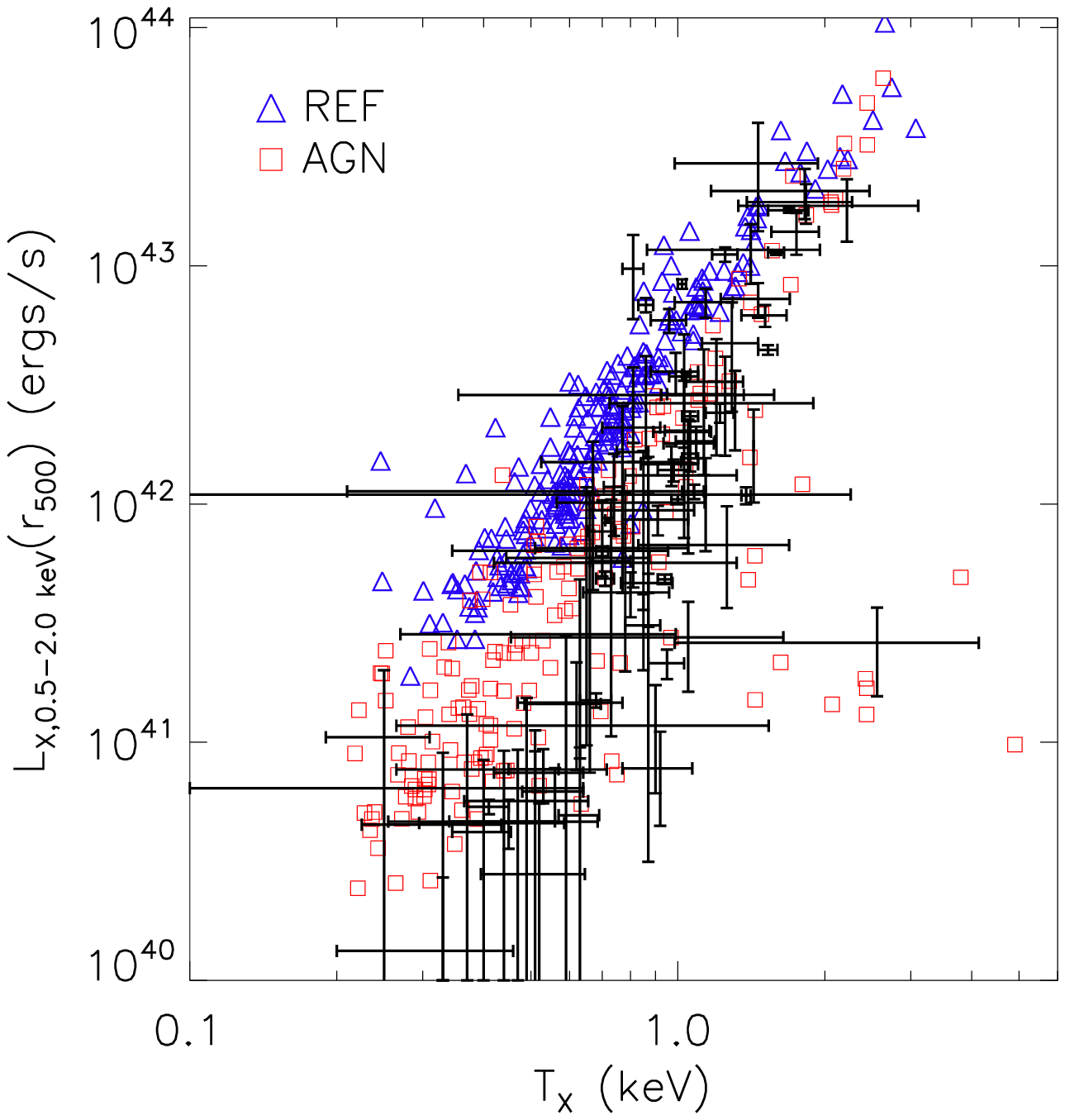}
\includegraphics[scale=0.55]{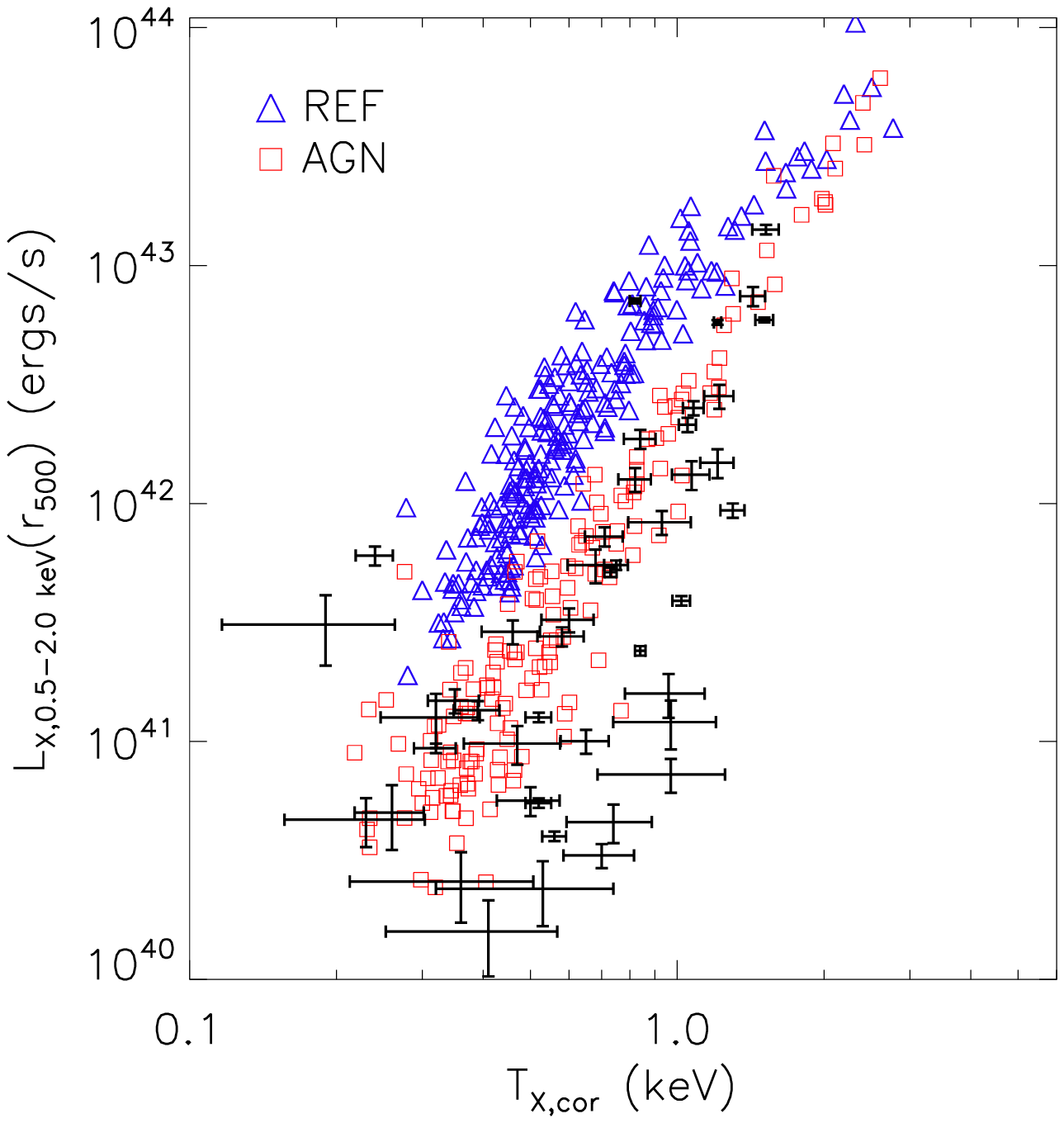}
\caption{X-ray luminosity-temperature relation.  {\it Left:}
X-ray luminosity (0.5-2.0 keV band) vs.\ uncorrected emission-weighted
temperature.  {\it Right:} X-ray luminosity (0.5-2.0 keV band) vs.\ cool core-corrected
emission-weighted temperature.
Points with error bars in the left hand panel represent the {\it ROSAT} measurements of Helsdon \& Ponman (2000) and Mulchaey et al.\ (2003) (uncorrected temperatures), while points with error bars in the
right hand panel represent the {\it ROSAT} measurements of
Osmond \& Ponman (2004) (corrected temperatures).  The {\it AGN} run yields a better match to both the corrected and uncorrected relationships, owing primarily to the reduced central gas densities in that run.    
 }
\label{fig:lt_relations}
\end{figure*}

It may at first sight be somewhat surprising that the {\it AGN} and {\it REF} runs 
have such similar mass-temperature relations.  This is primarily a consequence of the fact 
that the models produce fairly similar entropy distributions and, perhaps more 
importantly, that the X-ray temperatures have been `cool core-corrected'.  Cool core 
correction is a standard procedure designed to reduce the scatter in scaling 
relations between some observable and system mass, usually for the purposes of yielding 
more precise constraints on cosmological parameters derived from tests involving the use 
of cluster masses (e.g., the cluster mass function).  Typically, 
the correction involves excluding the central regions (the cool core, if one is present) when 
computing the mean system temperature\footnote{The term `cool core-correction' has also 
been applied to the procedure of calculating the luminosity after excising the emission from 
the central regions or from integrating a parametric model (e.g., 
a beta model) that has been fit to the X-ray surface brightness at large radii (typically 
beyond the cooling radius) and extrapolated inward.  However, when we apply it, we mean 
only the correction of the mean temperature, by computing over the projected radial range of $0.15-1.0 r_{500}$.}.  As we demonstrated in Section 3.1.2, the two models produce fairly different temperature profiles within 
the very central regions ($r \la 0.1 r_{500}$) of the groups.  Uncorrected scaling 
relations, therefore, are more useful for probing the astrophysics in the cores of groups 
and clusters.  This is illustrated more clearly below.

\subsubsection{Luminosity-temperature relation}

In Fig.\ 6 we plot the 0.5-2.0 keV X-ray luminosity vs.\ both the uncorrected (left 
panel) and cool core-corrected (right panel) emission-weighted temperatures.
For comparison, we 
show in the left panel the {\it ROSAT} measurements of Helsdon \& Ponman (2000) and Mulchaey et al.\ (2003) (temperatures uncorrected) and in the right panel the {\it ROSAT} measurements of Osmond \& Ponman (2004) (cool core-corrected temperatures).  

Focusing first on the left panel of Fig.\ 6, we see that the uncorrected $L_X-T_X$ relation from the 
{\it AGN} run is slightly steeper than that from the {\it REF} run.  This is a consequence of the reduced gas densities in the former.  Neither run, however, produces a relation that is as steep as the observed one, although the {\it AGN} run clearly performs better than the run with supernova feedback alone.   

Interestingly, the scatter in the {\it AGN} run is larger than that in the 
{\it REF} run, with the former having a small population of outliers (with $T_X \ga 1$ 
keV and $L_{X} \la 10^{42}$ ergs/s) that lie well off the mean relation.  
What is the origin of the increased scatter and the outliers in the {\it AGN} run?  

In the right hand panel of Fig.\ 6, we show the luminosity - cool core-corrected 
temperature relation.  We see here that the scatter in the {\it AGN} run has been reduced in general and also that there are no significant outliers off the mean relation.  This immediately indicates that both the increased scatter of the main population and outliers in the left panel are due to variations in the temperature of the gas within $0.15 r_{500}$, since the luminosity for the simulated groups is the same in both panels.  That feedback from AGN can increase the scatter in the luminosity-temperature relation through its effect on the mean temperature appears to be consistent with the observational findings of Croston et al.\ (2005).

The outliers in the left panel of Fig.\ 6 must have very hot, yet dense (i.e., dense enough to affect the mean system temperature) gas in their central 
regions.  It is likely that these systems have undergone very vigorous heating by their central supermassive BHs recently.  The temperature increase of gas heated by the central AGN is set to $10^8$ K in the current run, which is $\approx 
8.7$ keV.  If a sufficient number of particles have been heated, this would naturally 
explain the presence of the outliers.  However, the gas must have been heated quite 
recently (within the past $10^8$ yrs or so), so that the heated gas has not had sufficient time to rise buoyantly out of the central regions.  

Observationally, high temperature outliers appear to be rare (see data in left hand panel), which may signal a problem with the current model of AGN feedback.  It is 
possible that this problem, if real, may be simply resolved by {\it increasing} the 
temperature to which the gas is heated by the BH (to, say, $10^9$ K), so that the heated 
gas is so hot that it does not shine in soft X-rays, and therefore will not contribute 
strongly to the emission-weighted temperature (see discussion in McCarthy et al.\ 2008).  
Increasing the temperature to which the gas is heated should also result in 
longer periods of time between outbursts, which should further decrease the fraction of 
outliers at any given time.  The longer time between outbursts is required as more gas 
must be accreted by the BH in order for it to have sufficient energy to heat the gas to the higher temperature.
Alternatively, a change in the detailed implementation of the AGN feedback may also resolve this issue.  For example, Sijacki et al.\ (2008) implemented a mode of feedback which inflates bubbles containing a {\it non-thermal} relativistic population of cosmic rays.  It would be interesting to see whether or not such a model avoids producing outliers like the ones we observe in the present \owls {\it AGN} run.

A comparison to the observational data in the right panel of Fig.\ 6 shows that the {\it 
AGN} run matches the luminosity - cool core-corrected temperature relation relatively 
well, certainly much better than the run with supernova feedback only.  Since the cool 
core-corrected temperatures are similar for the two models at fixed system mass (Fig.\ 
5), the reason the {\it AGN} run matches the data must be that the X-ray luminosity 
has been reduced by a decrease in gas density owing to feedback from BHs.

Puchwein et al.\ (2008) and Fabjan et al.\ (2009) also compared their zoomed cosmological simulations to the observed luminosity-temperature relation of groups and clusters and likewise concluded that the addition of energy input from supermassive BHs results in a better match to the data, particularly for lower mass systems.  However, Fabjan et al.\ found this came at the expense of producing too high of a central entropy on the group scale.  We see no sign of this problem for the {\it AGN} run.  Improved agreement with the observed luminosity-temperature relation has also been found when BH feedback is incorporated into semi-analytic models (e.g., Bower et al.\ 2008) and hybrid methods (Short \& Thomas 2009).

\begin{figure*}
\includegraphics[scale=0.55]{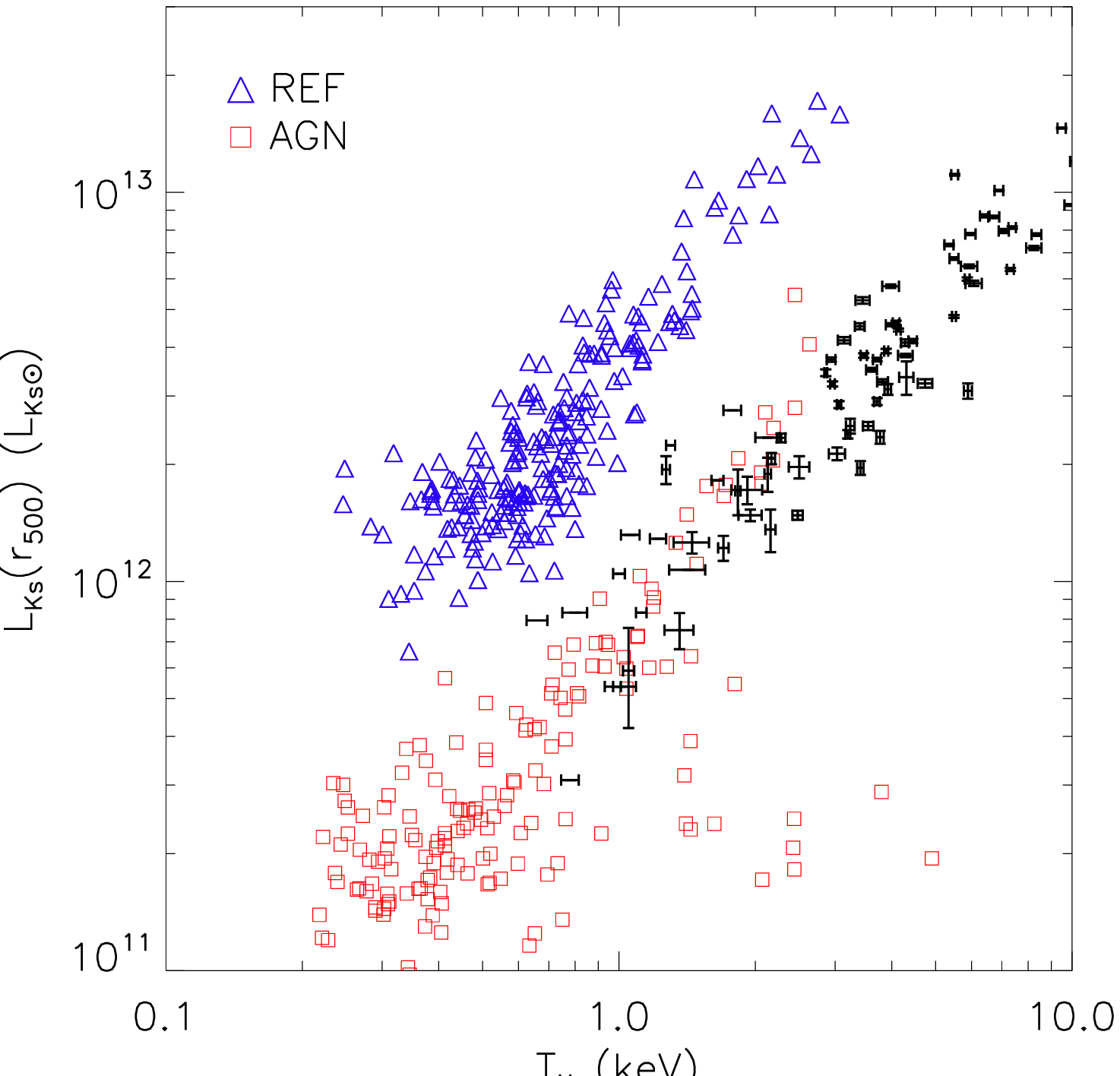}
\includegraphics[scale=0.55]{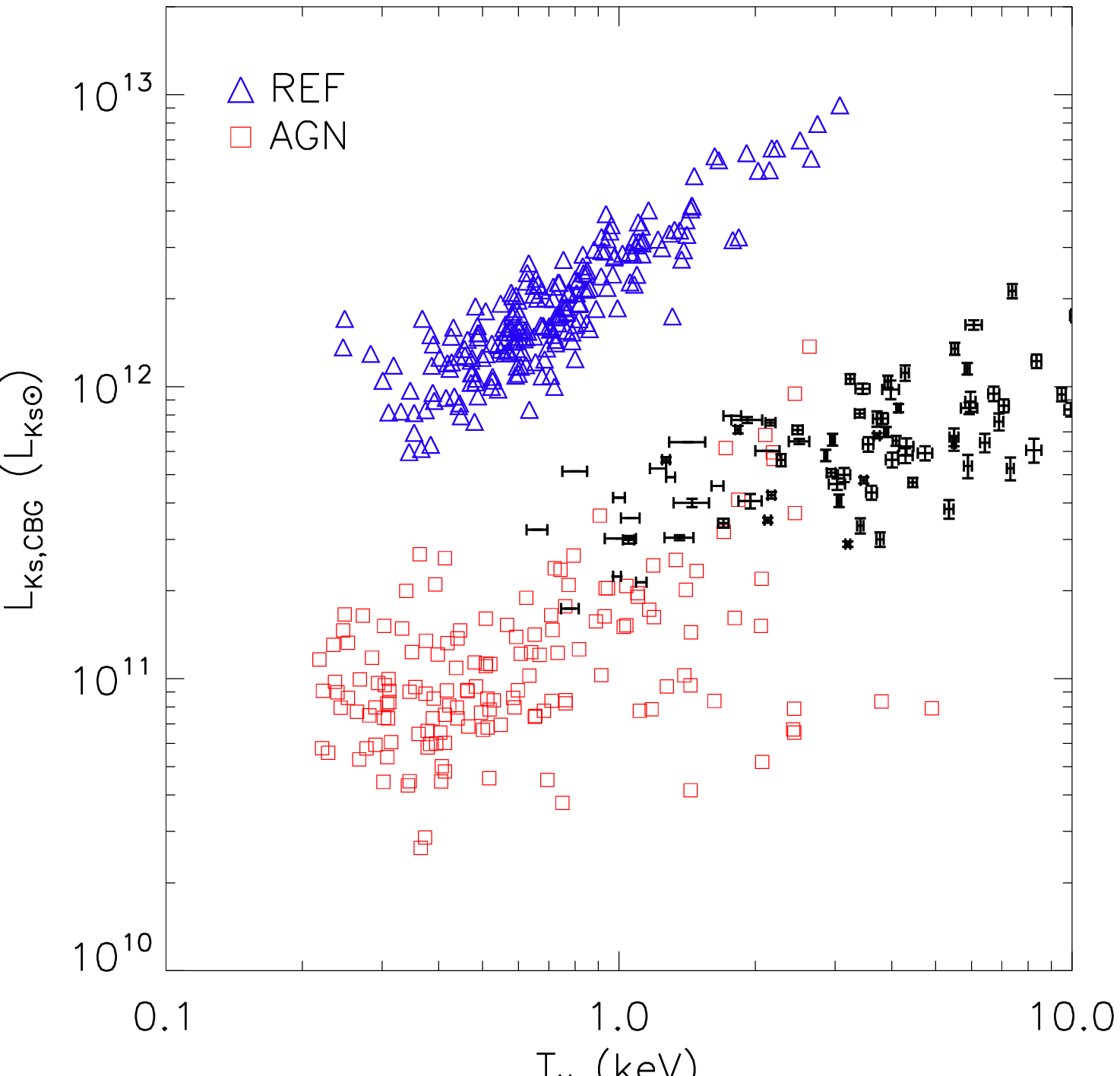} 
\caption{K-band luminosity within $r_{500}$ ({\it Left}) and of the CBG ({\it
Right}) vs.\ uncorrected emission-weighted temperature.  The points with error
bars on both luminosity and temperature represent the data of Lin \& Mohr (2004;
K-band luminosity) and Horner (2001; X-ray temperature).  The points with error 
bars only on the temperature represent the measurements of Rasmussen \&
Ponman (2009).  Feedback from supermassive BHs reduces the efficiency of star formation
by approximately a factor of 4, bringing the predicted $L_K-T_X$ relation into good
agreement with the observations.
 }
\label{fig:LK}
\end{figure*} 

\subsection{Formation of Stars}

\subsubsection{Group luminosity and luminosity of the CBG}

Galaxy groups and clusters are special objects in the universe, as they are the 
only objects for which it is presently possible to directly measure the total mass and 
distribution of both stars and hot gas (which dominate the baryonic mass budget)
out to a large fraction of the virial radius.  Over the past decade or so, cosmological 
hydrodynamic simulations have had varying degrees of success when it comes to matching 
the observed properties of the hot gas.  Consistently, however, it has proved very 
difficult to construct physical models that are able to match the stellar properties of 
groups and clusters.  In particular, observed groups and clusters show only a modest star 
formation efficiency of $\sim 10\%$ (i.e., only 10\% of the baryonic mass is in the form 
of stars; see, e.g., Balogh et al.\ 2008), whereas it is not uncommon for cosmological 
hydrodynamic simulations to yield much larger\footnote{We note here that an important factor 
in comparing different cosmological simulations is whether or not the simulations in 
question included radiative cooling losses due to metal lines.  As we will
show in a forthcoming paper, simply changing from primordial cooling rates to including metal-dependent radiative cooling as well, can 
increase the stellar fraction by a factor of 2 or more.} efficiencies of 30\%-40\%.  This 
general over-efficiency problem has been termed the `cooling crisis' of cosmological 
simulations (Balogh et al.\ 2001).

It is important to recognise that this cooling crisis appeared in, and still only 
strictly applies to, cosmological simulations that invoke feedback from supernovae winds 
(or no feedback at all) but not from supermassive BHs.  The recognition in recent years 
that heat input from supermassive BHs can significantly affect the properties of the gas, 
and therefore its ability to form stars, motivates us to re-examine the star formation 
efficiency of groups in cosmological simulations.

In Figure 7 we plot the K-band stellar luminosity within $r_{500}$ (left panel) and of 
the central brightest galaxy (CBG) (right panel) vs.\ the uncorrected emission-weighted mean 
X-ray temperature.  Because it is in the near-infrared, the K-band luminosity is sensitive 
mainly to older stellar populations and therefore should be a much better tracer of 
stellar mass than, say, optical luminosities in the B or V bands.  
Furthermore, it should be relatively unaffected by dust extinction.  Shown for 
comparison is the observational data of Lin \& Mohr (2004; {\it 2MASS} K-band 
luminosities) and Horner (2001; {\it ASCA} uncorrected X-ray temperatures).  The K-band luminosity of each star 
particle in the simulations is calculated using their formation times and metallicities and applying the simple stellar population models of 
Bruzual \& Charlot (2003), with no dust correction.  The CBG is identified by centroiding on mock K-band surface brightness maps of the simulated groups.  When quoting properties 
of the CBG below, we are referring to properties of the stars within 30 {\it h}$^{-1}$ kpc of the CBG 
centre.  For comparison, the typical effective radius of an observed CBG is 10-15 kpc 
(e.g., Shen et al.\ 2003).

From Fig.\ 7 it is immediately apparent that the {\it REF} model produces far too many 
stars with respect to the observations.  There are too many stars within $r_{500}$, by 
approximately a factor of 4, and too many within the CBG, by a similar factor.  This is 
the well known overcooling problem.  Inclusion of feedback from supermassive BHs, 
however, has a dramatic effect on the ability of the gas to form stars.  So much so that 
the star formation efficiency is lowered by almost exactly the factor that is required to 
match the data.  Note that the parameters of the AGN feedback were not tuned in any way to 
get this result (they were tuned to match the normalisation of the observed BH scaling relations; BS09).  Impressively, the slope of the observed relation is also reproduced 
relatively well by the simulations.  Note that the outliers in the $L_K-T_X$ relation of 
the {\it AGN} run are, as in Section 3.3.2, those systems which have anomalous central temperatures due 
to a very recent outburst - the scatter is much reduced when plotting vs.\ cool 
core-corrected temperatures, although no such temperatures are available for the Lin \& Mohr 
K-band CBG sample.

We point out that the fact that the run that incorporates energy input from supermassive BHs
results in CBGs with reasonable luminosities, while the run that does not over-predicts the stellar mass of the
CBG, agrees well with our expectations based on the analysis of the temperature profiles in Section 3.1.2. In particular, the temperature profile rises all the way into the centers of groups in the run that neglects AGN feedback, indicating that the central potential wells are much deeper in that run.  Therefore,
X-ray observations offer a probe independent of optical methods of the efficiency of star formation
in groups and clusters.

As a word of caution, the overlap (in terms of system mass) between the observations and simulations needs to be increased in order to be more definitive about our conclusions based on Fig.\ 7.  Observationally, this is difficult as obtaining temperature measurements for the hot gas in very low mass groups will likely require very long integration times and a very careful accounting of foreground and background X-ray sources.  The accuracy in the assignment of group galaxy membership also becomes much more important in the limit of low mass systems with few galaxies.  Alternatively, one can run simulations in larger periodic boxes to obtain more massive systems, for which observations already exist.  This has the disadvantage of focusing on relatively rare systems, so there is a worry about how general the conclusions obtained from comparing very massive simulated and observed systems are.

\begin{figure*}
\includegraphics[scale=0.55]{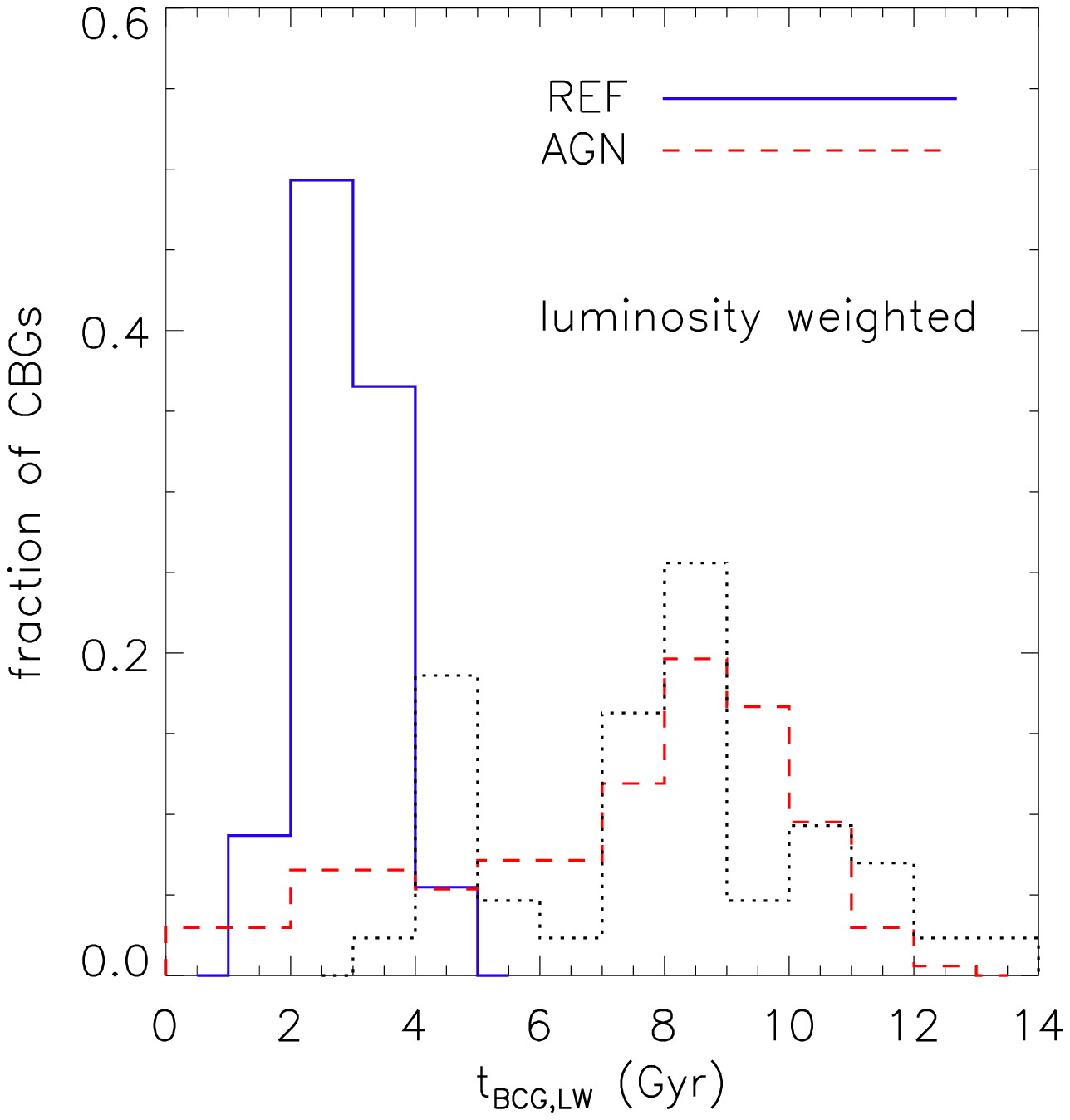}
\includegraphics[scale=0.55]{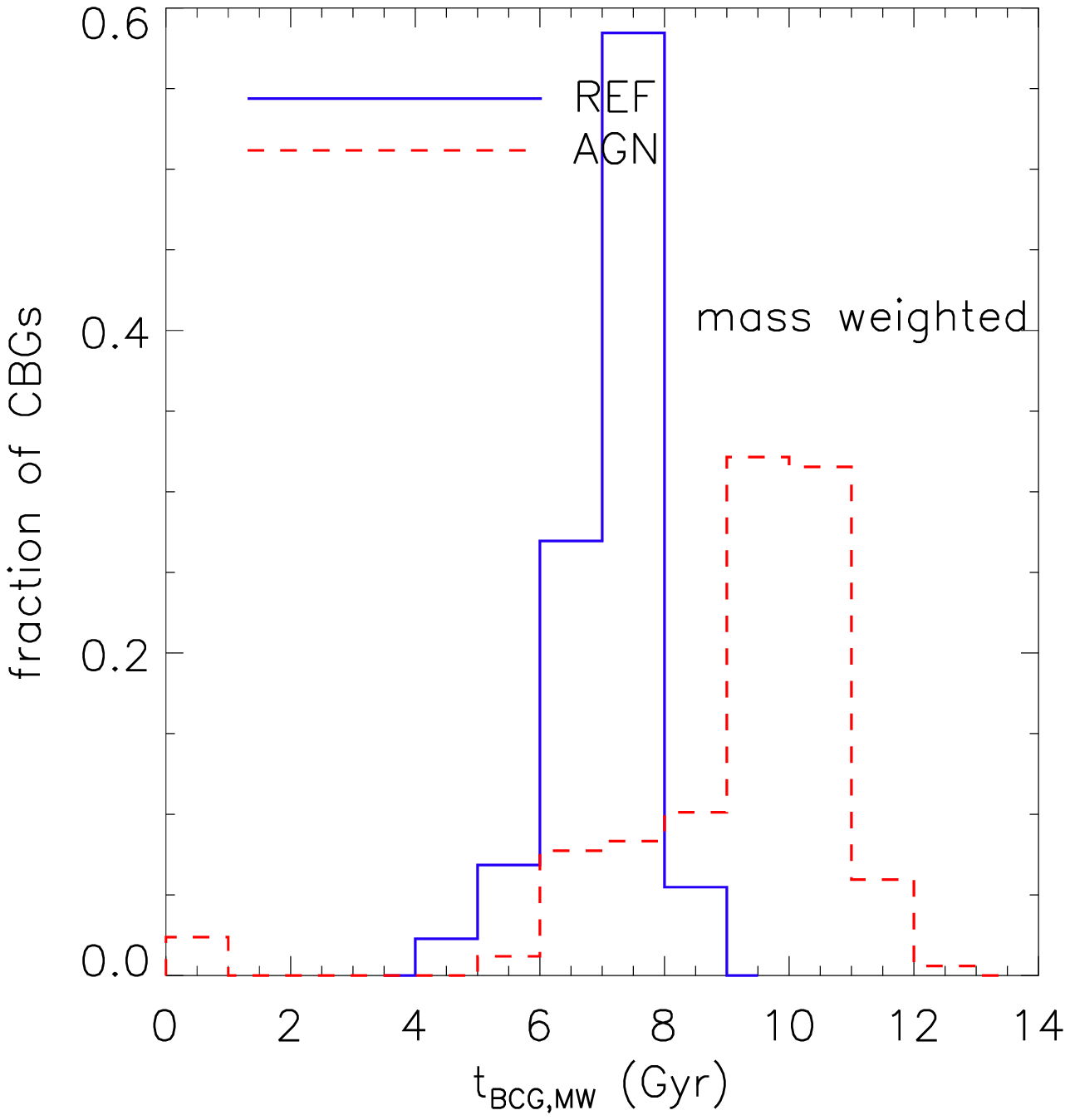}
\caption{B-band emission-weighted ({\it Left}) and mass-weighted ({\it Right}) stellar
age of the CBG.  Solid blue and dashed red histograms represent the {\it REF} and {\it AGN} \owls  
runs, respectively.  In the left panel, the dotted black histogram represents
observational data of Loubser et al.\ (2009).  Energy input from supermassive BHs significantly reduces
the star formation rates of CBGs at late times, yielding much higher (older) mean CBG ages in good agreement with observations.
}
\label{fig:stellarage}
\end{figure*}

In addition, it should also be kept in mind that the luminosities reported 
for the simulations represent the summation of all star particles within $r_{500}$, 
whereas the reported observed luminosities represent the summation of light above some 
surface brightness limit.  It is therefore possible that the present {\it AGN} run may 
have even yielded a slightly too {\it low} star formation efficiency.  There is currently 
a debate in the literature over what fraction of stellar mass/luminosity is in a diffuse, 
extended component (the intracluster light, or ICL for short), with estimates ranging 
from 10\%-50\% of the total light being in the ICL (Lin \& Mohr 2004; Murante et al.\ 2004; Zibetti 2005; 
Gonzalez et al.\ 2007; Balogh et al.\ 2008; McGee \& Balogh 2009).  The agreement with the central temperature profiles of observed groups (see Fig.\ 3) likely indicates that the fraction of stellar material in the central regions that is below the 2MASS surface brightness limit is small.  In any case, it is very clear that there is 
insufficient mass/light in the ICL to reconcile supernovae feedback-only simulations with 
observations.  More importantly, we have demonstrated that the inclusion of a 
physically-motivated model for feedback from supermassive BHs can significantly reduce the efficiency of cooling of gas destined to end up in groups (or group galaxies).  This is consistent with findings of several previous studies, including Sijacki et al.\ (2007), Fabjan et al.\ (2009), and Puchwein et al.\ (2008, 2010).

\subsubsection{Mean age of the CBG}

The simulations should not only match the total present-day stellar mass, but 
they must also reproduce the star formation {\it histories} of these systems, including 
their present-day star formation rates.  Historically, supernovae 
feedback-only simulations have not only failed to yield reasonable stellar luminosity and mass functions, 
they have also failed to reproduce the colours of observed CBGs, with the simulated 
galaxies being too blue with respect to the observed systems.  This is an 
indication that recent star formation is too efficient in these simulations.  Another 
way of expressing the same thing is that the mean stellar age of the CBG is too young 
compared to the observed ages.  We now examine whether including feedback from supermassive 
BHs affects the mean stellar age of the CBG.

In Fig.\ 8 we show histograms of the B-band emission-weighted (left panel) and 
mass-weighted (right panel) CBG age for the {\it REF} and {\it AGN} \owls runs.  Also shown, in the left 
panel, is the data of Loubser et al.\ (2009).  These authors derived the stellar ages of 
approximately 50 CBGs in a sample of nearby groups and clusters from long slit spectra using the 
stellar population synthesis models of Maraston et al.\ (2003, 2004) with the Lick system 
of absorption line indices. (The advantage of using the absorption line indices is that they should not be affected by dust.)  The 
derived quantities are obviously emission-weighted and because they are in the optical a 
comparison to the B-band weighted simulation ages is appropriate (we have also compared to 
the V-band weighted simulation ages and find very similar results).

Focusing first on a comparison between the models, we see that when the stellar ages are 
luminosity weighted there is a large difference in the mean age of the CBG between the 
two models, with the peak of the distribution being between $\approx2-3$ Gyr for the {\it 
REF} run and between $\approx8-9$ Gyr for the {\it AGN} run.  The mass-weighted 
distributions are more similar and indicate a generally old population of stars in both runs.  This is expected since most of the stellar mass forms at $z \sim 1-3$ (corresponding to the peak of the star formation rate density of the universe), even in simulations with inefficient feedback such as the {\it REF} run.  These results indicate that the B-band luminosity 
weighted ages are very sensitive to recent episodes of star formation and that the star formation rate of CBGs in the {\it REF} run are presently (or have very recently been) much higher than those 
of CBGs in the simulation that includes AGN feedback.

A comparison to the data of Loubser et al.\ (2009) confirms our expectation that the {\it 
REF} model yields mean stellar ages that are far too young compared to real CBGs.  While 
we should expect an improvement by including feedback from supermassive BHs, 
agreement with the data is not necessarily automatic.  It will obviously depend on when 
and where (that is, in what progenitors) the BHs heated the gas, and one can envisage 
not unreasonable scenarios where the integral of the star formation rate history is 
reproduced but not the detailed history itself.  In spite of this, the {\it AGN} run 
reproduces the distribution of stellar ages of observed CBGs well, especially the peak of 
the distribution.  The large scatter is also retrieved relatively well.  

On a qualitative level, our results confirm the findings of De Lucia et al.\ (2006), who used a semi-analytic model of galaxy formation to show that the inclusion of a prescription of AGN feedback (see Croton et al.\ 2006) results in much older formation ages for massive elliptical galaxies.

\begin{figure}
\includegraphics[scale=0.55]{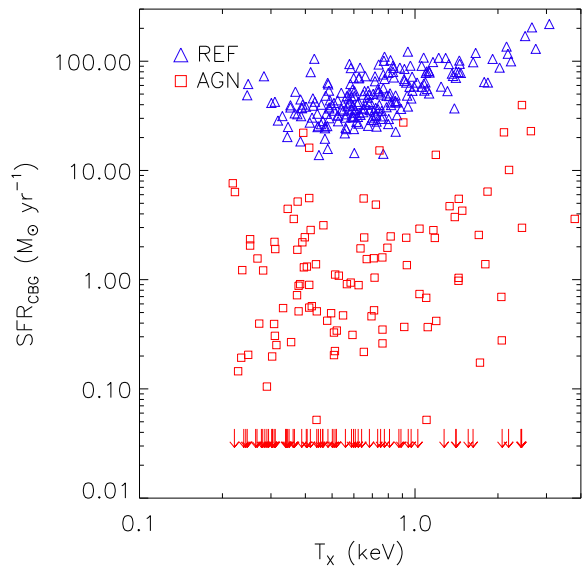}
\caption{
Present-day ($z=0$) star formation rate of the central brightest galaxy (CBG) vs.\
uncorrected emission-weighted temperature.  The red arrows denote upper limits on the star formation rate for CBGs in the {\it AGN} run with no star forming gas particles.  The upper limit is the minimum resolvable SFR in the simulations, corresponding to the SFR of a single particle with a density equal to the threshold for star formation.  AGN feedback dramatically reduces the present-day star formation rates of CBGs, shutting off star formation altogether in most systems.
 }
\label{fig:centralsfr}
\end{figure}

\subsubsection{Present-day star formation rate of the CBG}

We now examine the present-day star forming properties of the CBGs in the two runs.  It is important to note that it is entirely plausible that a model could match the K-band luminosity of the CBG without matching the present-day star formation rate.   This is because most of the stellar mass was not formed recently out of a cooling flow, even in models where there is no feedback at all and cooling flows operate uninhibited at the present day.  Most of the stellar mass in groups and clusters was formed in galaxies at high redshift ($z \sim 1-3$).  Therefore shutting down a cooling flow
today will not help with the overcooling problem of cosmological simulations if excessive cooling is not also shut down in galaxies at high redshift.  By the same token, getting the correct star formation rate of galaxies at high redshift does not guarantee prevention of cooling flows today.

We show in Fig.\ 9 the present-day star formation rate of gas in the CBG vs.\ 
uncorrected X-ray temperature.  A tight trend between these quantities exists for the {\it 
REF} run, with the SFR of the CBG increasing with system temperature (mass).  At a temperature 
of 1 keV (M$_{500} \approx 3\times10^{13}$ M$_\odot$), CBGs in the {\it REF} groups are 
forming stars at a rate of $\sim50-100$ M$_\odot$ yr$^{-1}$.  By contrast, when feedback 
from 
supermassive BHs is included, the star formation rates are drastically reduced and 
in most cases star formation is shut off altogether.

What are the present-day star formation rates of real CBGs?  This is actually not a
straightforward question to answer.  One does not directly 
measure the star formation rate, but instead uses observables (such as infrared, 
H-alpha, and radio luminosities) as proxies for the star formation rate.  However, these 
observables do not correlate perfectly with the star formation rate and  
systematics such as dust or aperture corrections can be very important\footnote{Hicks et al.\ (2009), for example, find very large discrepancies (up to an order of magnitude) between UV and IR-inferred SFRs for CBGs in a sample of 16 cool core clusters.}.  Also, one must 
take care to account for differences in the assumed stellar IMF.  Thus, accurate 
derivation of absolute star formation rates for CBGs is non-trivial.  One 
might expect, though, that the observations can be used to robustly determine the {\it 
fraction} of CBGs that are forming stars at a reasonable (detectable) rate.

The fraction of CBGs forming stars has been shown to depend 
somewhat on sample selection criteria, as there is a known correlation between the cooling rate of 
the hot gas and whether or not there is (or has recently been) star formation in the 
central object (e.g., Crawford et al.\ 1999; Edwards et al.\ 2007; Bildfell et al.\ 
2008; Rafferty et al.\ 2008; Wang et al.\ 2009).  Thus, a sample containing a large fraction of cool core groups and clusters (as 
one might expect in the case of X-ray flux-limited samples) should be expected to host 
a larger fraction of star forming CBGs than a sample that contains a small fraction of such systems.  For example, Crawford et al.\ (1999) found that for a sample of 256 
dominant galaxies in 215 nearby X-ray-selected groups and clusters, approximately 27 per 
cent have emission line spectra.  More recently, Edwards et al.\ (2007) examined the 
optical spectra of CBGs in two separate samples; an X-ray-selected sample based on the 
NOAO Fundamental Plane Survey and an optically-selected sample based on the SDSS DR3 
survey.  Edwards et al.\ found that only $\sim13\%$ of the optically-selected CBGs 
contained line emission, whereas $\sim20\%$ of the X-ray-selected CBGs contained 
detectable line emission.  A similar trend, based on larger optical and X-ray samples, has been reported recently by Wang et al.\ (2009).  While Edwards et al.\ and Wang et al.\ did not quantify what fraction 
of their optically-selected groups/clusters contained cool cores (as many of their optical systems have no existing pointed X-ray observations), there is no obvious reason why these systems should be biased for or biased against in an optical selection.

\begin{figure*}
\includegraphics[scale=0.55]{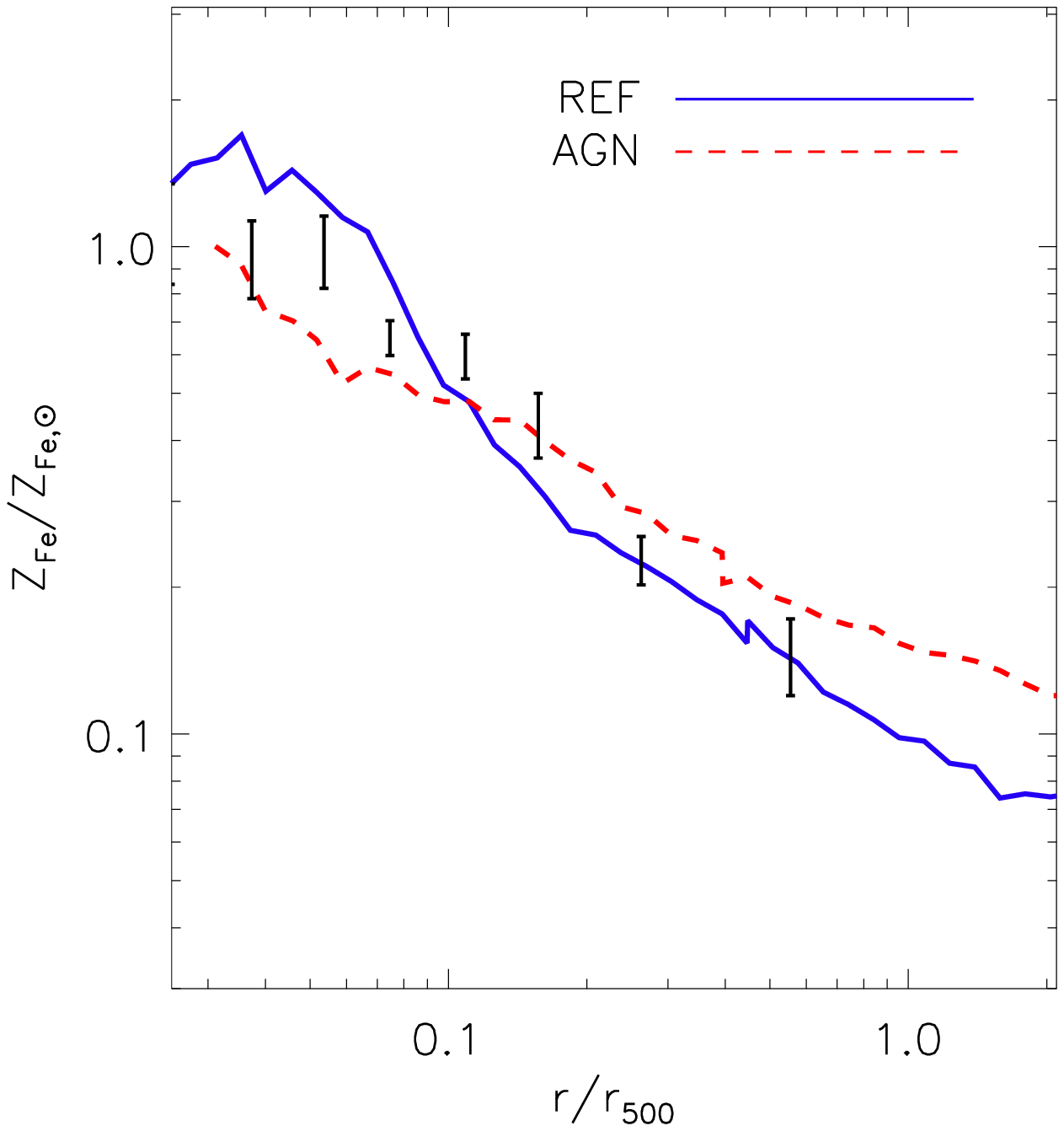}
\includegraphics[scale=0.55]{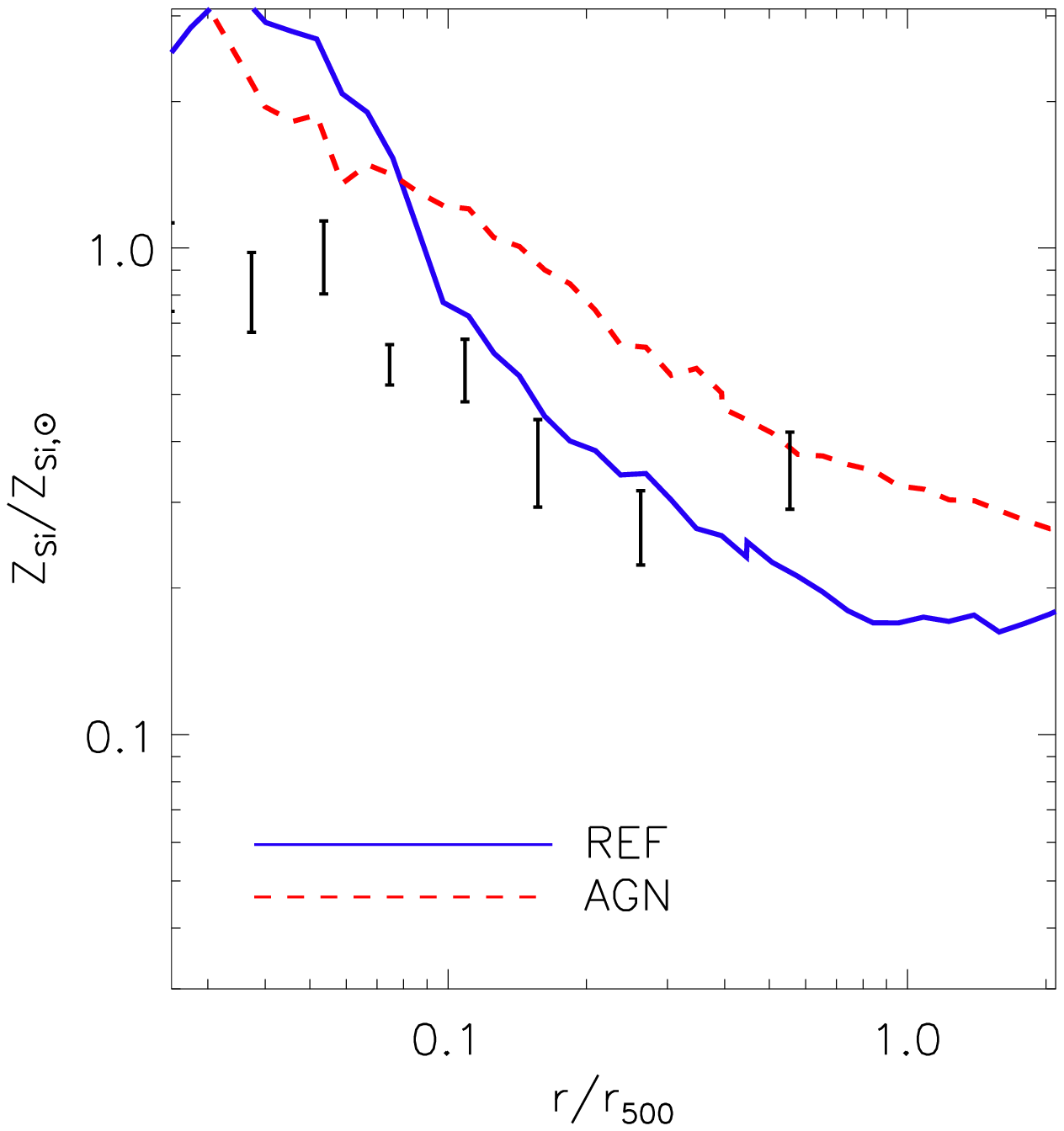}
\caption{
Median emission-weighted 3D iron ({\it Left}) and silicon ({\it Right})
abundance profiles for the {\it REF} and {\it AGN} \owls runs.  Error bars represent the
binned averages of the {\it Chandra} data of Rasmussen \& Ponman (2009).
Both the
observational and theoretical radial profiles are normalised using the solar abundances of Asplund et al.\ (2005).  Both models reproduce the iron data relatively well.  The shape of the silicon profile agrees better with the run including AGN, but the predicted abundance may be slightly too high. 
} \label{fig:gasmetals1}
\end{figure*}

The resulting fraction of CBGs that have robust line emission therefore likely 
lies between $10\%-30\%$.  Assuming this line emission is due to ongoing star formation 
(as opposed to, e.g., AGN), would imply that this fraction of CBGs are forming stars at a 
rate of at least a few solar masses per year, assuming a Chabrier IMF (M.\ Balogh, 
priv.\ communication).  However, this likely represents an upper limit, as some of the 
optical line emission will not have originated from recent star formation.  For example, O'Dea 
et al.\ (2008) examined a sample of 62 CBGs {\it with optical line emission} in the 
mid-infrared with the {\it Spitzer} Space Telescope and found that only half of the 
systems showed evidence for a detectable excess of mid-IR emission.  Thus, the fraction 
of CBGs with ongoing star formation could actually be closer to only $5\%-15\%$.   By 
contrast, 100\% of the CBGs in the \owls {\it REF} run are forming stars at a rate that 
would be detectable with present observations.  When feedback from AGN is included, we 
find the fraction of CBGs that are forming stars at a rate of $>3$ M$_\odot$ yr$^{-1}$ is 
reduced to only 14\%, which is comparable to the observed fraction.

Interestingly, for those systems that {\it are} forming stars in the {\it AGN} run, there 
is a large degree of scatter in the SFR.  This may be linked with the large degree of 
scatter seen in the mean ages of the CBGs (Fig.\ 8) and in the central entropy of the hot 
gas (left panel of Fig.\ 2).  In a future paper we will examine the physical origin of the scatter in the star forming properties of CBGs in simulations with AGN feedback.

\subsection{Metallicity}

The metallicity of gas plays a crucial factor in its ability to cool and form stars.  The 
cooling rate obviously also has implications for the thermodynamic properties of the hot 
gas that has not yet turned into stars, particularly on the galaxy group scale where a large fraction of the X-ray emission 
originates from line emission.  In order to match the 
thermodynamic properties of the hot gas in groups in detail with a physical model, one should therefore also match the metallicity distribution of that gas.  It is therefore important that simulations allow for  
metal-dependent cooling losses and obviously for the physical mechanisms that are 
responsible for producing and getting metals out into the gas.  

As described in Wiersma 
et al.\ (2009a), the \owls simulations compute cooling rates on an 
element-by-element basis for the 11 most important species for 
cooling.  The timed release of these elements (into the gas surrounding star particles) 
by Type II and Type Ia SNe, as well as from intermediate mass stars (`AGB' stars), is 
tracked during the simulations (Wiersma et al.\ 2009b).  Whether or not these metals will 
then get distributed through the intragroup medium will depend on whether the 
metal-enriched gas cools quickly and gets locked back up in stars or, if not, if 
feedback and/or dynamical processes associated with the group environment are sufficient to move metals from haloes of individual galaxies into the intragroup medium.  Thus, the gas- 
and stellar-phase metallicities of groups provide an important diagnostic of the 
efficiency of star formation, feedback, and dynamical processes (such as ram pressure and 
tidal stripping) in groups.  We now examine how well the \owls {\it REF} and {\it AGN} simulations are able to 
reproduce the observed metal content of groups.

Before proceeding, it is important to note that nucleosynthesis yields and SN Ia rates are both uncertain at the factor of $\sim2$ level, even for a fixed stellar initial mass function (see Wiersma et al.\ 2009b).  Therefore, at least at present, we argue that more weight should be placed on the successes/failures of simulations in reproducing the thermodynamic state of the gas and the overall baryonic mass fractions (discussed in Sections 3.1-3.4).  Metal abundances can only be expected to match the observations to within a factor of $\sim 2$.

\subsubsection{Gas-phase metallicity}

In Fig.\ 10 we plot the median emission-weighted iron and silicon profiles for the {\it 
REF} and {\it AGN} runs.  For comparison, we plot the recent {\it Chandra}-derived results of Rasmussen \& Ponman 
(2009) for a sample of 15 bright galaxy groups.  

Focusing first on the profiles, we see that both \owls runs produce similar large-scale 
metallicity gradients, with the iron (silicon) abundance reaching solar values (several 
times solar) at the centre and declining to a tenth (a few tenths) solar by 
$r_{500}$.  In fact, large-scale metallicity gradients such as these are a generic 
feature of all of the \owls runs.  In addition, we note that the profiles presented 
here appear to be very similar in shape to those found in the zoomed cosmological 
simulations of massive clusters in Tornatore et al.\ (2007) and Borgani et al.\ (2008).  

\begin{figure}
\includegraphics[scale=0.55]{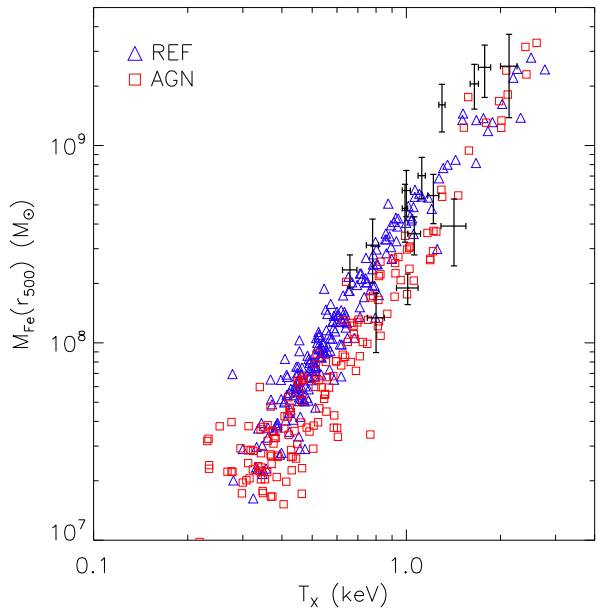}
\caption{
Total mass of hot gas-phase iron as a function of cool core-corrected emission-weighted X-ray temperature. Error bars represent the measurements of Rasmussen \& Ponman (2009).  Both models agree with the data.
} \label{fig:gasmetals2}
\end{figure}

The shape and normalisation of the simulated iron profiles are remarkably similar to 
the binned average profile found by Rasmussen \& Ponman (2009).  In terms of cooling of hot gas, iron is the most important species.  It is therefore fortunate that the simulations reproduce the gas-phase iron abundance, given the uncertainty in the nucleosynthesis yields and SN Ia rates.  Both simulations appear to produce groups that are over-abundant in silicon.  The shape of the observed silicon profile, at least in the central regions, is more similar to that of the {\it AGN} run, so an adjustment of the assumed Si yields would improve the fit to the data.  One might naively expect that adjusting the yields could have important consequences for the thermodynamic state of the gas, and this would indeed be the case for iron.  However, silicon is a far less efficient coolant than iron for the temperature range relevant for this study (see Fig.\ 6 of Wiersma et al.\ 2009a).  Therefore, a change in the Si abundance would likely only have a very minor effect on the resulting gas properties.

The result that including feedback from supermassive BHs gives rise to ICM metallicity profiles whose shape is in better agreement with observed profiles, at least in the central regions, was also obtained recently by Fabjan et al.\ (2009).  On the other hand, neither the {\it REF} nor the {\it AGN} models reproduce the shape of the observed silicon profile at large radii.  Like Fabjan et al.\ we find that the Si/Fe ratio is nearly flat as a function of radius from roughly $0.2 r_{500}$ out to $r_{500}$, whereas observationally Rasmussen \& Ponman (2009) find a steep rise in Si/Fe beyond about $0.2 r_{500}$, which is due to the flattening of the Si profile at large radii.  However, observationally-inferred silicon abundances in the outer regions of groups are still controversial.  As noted by Fabjan et al.\ (2009), several other recent observational studies based on {\it Suzaku} data (e.g., Sato et al.\ 2008, 2009) find no evidence for such a steep rise in Si/Fe at large radii (although note that typically these studies have been limited to $\approx 0.3 r_{500}$).  If the results of Rasmussen \& Ponman (2009) are confirmed, this will signal a problem with the current models of AGN feedback that may not be rectified by a simple change of the adopted yields and/or SN Ia rates.

Instead of stacking the groups to obtain a single median profile, we plot in Fig.\ 11 the total mass of iron within $r_{500}$ vs.\ system temperature (mass).  When computing the 
total metal mass for the simulated groups, we sum only metals 
attached to gas particles that are hot, $> 10^5$ K, and therefore X-ray-emitting.
In accordance with the profiles results, the total mass of iron in the hot gas phase is in 
excellent agreement with the observations.  
We note that we have also computed the total mass of silicon within $r_{500}$ and compared it with the measurements of Rasmussen \& Ponman (2009) and find similarly good agreement.  This may seem surprising, given the profiles plotted in Fig.\ 10, but note that most of the metal mass is located at large radii, beyond $\sim 0.4 r_{500}$, where the {\it AGN} run reproduces the observed Si abundance relatively well.  However, this agreement may be misleading, as the measured Si abundances are most uncertain at large radii and, as already noted, the absolute abundances in the simulations are uncertain at a factor of 2 level owing to uncertainties in the adopted nucleosynthesis yields and SNIa rates.

\begin{figure}
\includegraphics[scale=0.55]{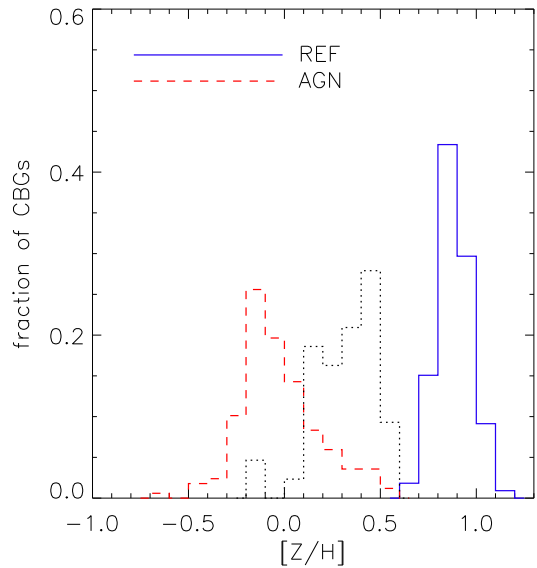}
\caption{B-band emission-weighted stellar-phase metallicity of the CBG, presented in logarithmic units and relative to the Sun.  Solid blue
and dashed red histograms represent the {\it REF} and {\it AGN} \owls runs, respectively.  The dotted
black histogram represents observational data of Loubser et al.\ (2009).  We adopt the
solar abundances of Asplund et al.\ (2005) in normalising the simulations.}
\label{fig:stellarmetals}
\end{figure}

\subsubsection{Stellar metallicity}

It is a curious fact that the {\it REF} and {\it AGN} runs produce such similar gas phase 
metal masses (Figs.\ 10-11), given that they have very different total stellar masses (Fig.\ 7).  This 
may seem to be a contradiction at first, but in fact the explanation for this behaviour is 
simple: the mass of metals locked up in stars is very different in the two runs.  In 
Fig.\ 12 we show the distribution of B-band emission-weighted CBG metallicities for the 
two \owls runs.  Also shown for comparison are the observational determinations of Loubser 
et al.\ (2009) for a sample of approximately 50 nearby CBGs in groups and clusters.

A comparison between the two simulations shows nearly an order of magnitude difference in the peaks of 
the two distributions, with the CBGs in {\it REF} run being far more metal rich than those in 
the {\it AGN} run.  The effect of feedback from supermassive BHs is therefore not 
only to quench the overall star formation rate, but also to prevent the metals that are 
produced from being locked back up in future generations of stars.  The BHs heat local gas causing it to expand out of the dense, star forming-regions and thereby raise the fraction of the total metal mass that is in the gaseous phase.  The efficient ejection of metals from star forming-regions is the mechanism by which the {\it AGN} run yields similar gas-phase abundances to the {\it REF} run, in spite of it having a stellar mass fraction that is $\approx 4$ times smaller than the latter run.  The BH heating also leads to increased mixing and a more uniform
metal distribution.  In this respect, our results agree, at least qualitatively, 
with those of Sijacki et al.\ (2007) and Fabjan et al.\ (2009).

Loubser et al.\ (2009) have recently measured the emission-weighted stellar metallicity 
of a large sample of CBGs.  It should be noted that the derived abundances 
are not fully self-consistent in terms of the assumed solar abundance, as the isochrones 
adopted by the Maraston et al.\ models (from Cassisi et al.\ 1998) assume $Z_\odot = 0.02$, 
whereas the stellar library is based on the solar neighbourhood and the metallicities of 
individual stars are given relative to the Sun (P. Sanchez-Blazquez, priv.\ communication).  
For the simulated CBGs we assume the 
up-to-date solar abundances of Asplund et al.\ (2005), which gives $Z_\odot = 0.012$.  

A 
comparison of the simulated CBGs to the data of Loubser et al.\ (2009) reveals that 
neither run performs particularly well compared to the data, with the {\it REF} run 
producing too metal rich and the {\it AGN} run producing too metal poor stars.  If we 
were to adopt, instead of the Asplund et al.\ (2005) abundances, a solar metallicity of 
$Z_\odot = 0.02$, this would shift the models to lower 
metallicities (in solar units), making the discrepancy larger for the {\it AGN} run and 
smaller for the {\it REF}.  However, even with this shift the {\it REF} model does 
not provide a good match to the data.  We have also compared the simulations 
to the [Fe/H] and [Mg/H] measurements of Carretero et al.\ (2007) for a sample of 27 
elliptical galaxies in 4 nearby massive clusters and found very similar results.

The apparent discrepancy may indicate that the adopted yields are not correct in detail and/or that the metal ejection resulting from supermassive BH feedback 
in the current simulation may be too efficient.  It is presently unclear what the specific 
cause of the discrepancy is.  If the problem is related to over-efficient metal ejection, we expect that rectifying this problem would not significantly affect the excellent match 
to the gas phase metallicity, as long as the match to the observed stellar content is maintained.  The reason for 
this is that only a small fraction of the baryons in observed groups are in 
the form of stars.  Thus, only a relatively small amount of metals would need to be removed 
from the intragroup medium to raise the stellar metallicity to the observed level.

\section{Discussion \& Conclusions}

We have analysed two high resolution cosmological hydrodynamic simulations (the {\it REF} and {\it AGN} runs) from the \owls project.  While both include galactic winds driven by supernovae, only one of the models includes feedback from accreting BHs (the {\it AGN} run).  These runs are representative of the \owls runs with supernovae feedback only and runs with AGN feedback, respectively, in terms of the galaxy groups properties they produce (Paper III).  We compared the runs to a wide range of observations in order to ascertain whether or not 
including BH feedback in cosmological simulations yields a more realistic population of 
galaxy groups.  This comparison has yielded the following findings:

\begin{itemize}
\item{The two models yield similar median entropy distributions for the gas and reproduce the 
`excess entropy' of observed groups with respect to the self-similar scaling. However, there is a much larger 
degree of scatter in the central entropies of groups in the run that includes BH feedback.  
This, in turn, gives rise to a larger scatter in the luminosity-temperature relation of this model.}
\item{The two models produce similar temperature profiles beyond $0.1 r_{500}$, but within this radius the temperature of groups in the {\it REF} model continue to rise, in discord with the observations.  This is due to the excessive accumulation of cold baryons in the centres of groups in the absence of AGN feedback, which significantly deepens the central potential well leading to compressional heating of the gas.}
\item{Supernovae-driven winds alone do not inject sufficient energy to reproduce the observed low 
gas mass fractions of galaxy groups.  The additional energy input from BHs is, however, 
sufficient to reduce the gas mass fractions of galaxy groups to the observed level and 
also reproduces the steep increase in the gas fraction with system temperature.}
\item{Both models match the observed scaling between mass and 
cool-corrected temperature relatively well.  This is because beyond the central regions the gravitational potential wells are similar in both models, owing to the dominance of dark matter at large radii.}
\item{The reduced gas density in the groups that have been subjected to BH feedback 
results in lower X-ray luminosities at fixed halo mass/mean temperature.  The resulting 
luminosity - cool core-corrected temperature relation is in good agreement with observations.  When the temperatures are not cool core-corrected, the scatter in the L-T relation 
is increased significantly for the {\it AGN} run, with several outliers lying well off the 
mean relation.  This is due to recent large episodes of BH heating.  Such outliers appear to be very rare in nature.  We speculate 
that if the heating temperature were increased to a higher level than $10^8$ K the scatter 
would be reduced, as the heated particles would cease to emit most of their radiation in 
soft X-rays.  Also, the length of time between outbursts would likely increase, as more 
mass will have to be accreted by the black hole in order to have sufficient energy to heat 
the gas to a higher temperature.}
\item{The {\it AGN} run yields stellar K-band luminosities of both overall group and the CBG that are in good agreement with observations.  The inclusion of feedback from BHs also yields the correct emission-weighted mean age of CBGs (as well as the scatter in age) and the fraction of CBGs that are forming stars at the present day ($\sim 15\%$).  In contrast, the 
simulation with supernova feedback alone produces CBGs that are far too massive and young (with very 
little scatter in mean age) and all of the CBGs in this run have significant present-day star formation rates.}
%\item{Interestingly, for those CBGs that are forming stars in the run with BH feedback, 
%the range in SFRs spans more than two orders of magnitude.  This large scatter may be 
%linked to the time since the last outburst and/or the large scatter seen in the central 
%entropy of the gas in these groups (see Fig.\ 2).  We are presently investigating the 
%origin of the scatter in the CBG SFR.}
\item{Both models yield total hot gas-phase iron  
masses that agree with observed masses and both yield large-scale gradients that are similar to the observed profiles.
However, while the simulations reproduce the 
observed slope and normalisation of the iron abundance profile, both appear to be over-abundant in silicon at small radii.  The discrepancy is, however, not greater than a factor of two and therefore lies within the current uncertainty in nucleosynthesis yields, so at present this is not a significant problem for the simulations.}
\item{Neither simulation yields the correct metallicity of stars, with the simulation with 
BH feedback producing too metal poor stars while the simulation without BH feedback 
produces too metal rich stars.  This may signal that the present implementation of AGN 
feedback is too efficient in expelling metals from star forming regions and/or that the adopted yields are incorrect in detail.  If it is the former, we note that since the mass 
in metals in the hot gas phase is much larger than that locked up in stars, only a 
relatively small fraction of metals would be needed to be removed from the intragroup 
medium to resolve this problem.}
\end{itemize}

Based on the above, the inclusion of feedback from supermassive black holes significantly improves the ability of our cosmological hydrodynamical simulations to yield a realistic population of 
galaxy groups.  This agrees qualitatively with the findings of Sijacki et al.\ (2007), Puchwein et al.\ (2008, 2010), and Fabjan et al.\ (2009), who implemented BH growth and feedback in cosmological zoomed simulations.  We note, however, that our conclusions are based on a comparison of a much larger sample of simulated groups to a wider range of observations.  In addition, the \owls implementation of BH feedback appears to have yielded better matches to several different key observables than those found in some of these previous studies.  For example, Fabjan et al.\ (2009) and Puchwein et al.\ (2008, 2010) report stellar mass fractions within $r_{500}$ that are a factor of $\sim 2$ higher than observed and what we obtain from the {\it AGN} model.  In addition, Fabjan et al.\ (2009) find their galaxy groups have too high entropies at $r_{2500}$.  Given the many differences in the detailed implementations of cooling, star formation, chemodynamics, and feedback in the \owls simulations and those which are based upon the model of SDH05, it is not trivial to determine what physics is most responsible for the better match to the data.  One possibility is the simulations of Sijacki et al.\ (2007) and Puchwein et al.\ (2008) did not include metal-line cooling which is expected to be important.

We have highlighted a number of areas where the current model of {\it AGN} feedback could be improved, including reproducing the scatter in the (uncorrected) luminosity-temperature relation and exploring whether or not simple adjustments in the adopted SNe yields will allow the simulations to better reproduce the gas-phase silicon abundance and stellar metallicities.  We leave this for future work.

Recently, Dav{\'e} et al.\ (2008) argued that it is possible to reproduce the entropy and metallicity distributions of the hot gas in groups with simulations that invoke galactic winds with scalings similar to those expected for outflows driven by radiation pressure from massive stars on dust grains, but which do not include AGN feedback.  While it is not straightforward to directly compare the \owls simulations with those of Dav{\'e} et al.\ (as there are differences in the implementations of  star formation, chemodynamics, cooling, and feedback), some of their results do not appear to be inconsistent with ours, as we have also demonstrated that models without AGN feedback can reproduce these quantities.  However, their model also appears to yield stellar mass fractions in reasonable agreement with the observations, whereas the \owls {\it REF} run does not.  Plausible reasons for this difference are that the energy injected in the momentum-driven wind implementation of Dav{\'e} et al.\ exceeds that available from supernovae (see Schaye et al.\ 2009). %and that their calculations neglect losses due to drag forces in the ISM.  
Thus, one must invoke alternative energy sources, namely high energy photons exerting pressure on dust grains, to motivate this model. 
%[but see also Haas et al.\ (in prep) who argue that Dav{\'e} et al.\ inject more momentum then is actually available from radiation from a simple stellar population, which represents the maximum momentum one can extract from stars.  Thus, as implemented, the momentum-driven winds cannot really be motivated based on photons coupling to dust either.]
By contrast, supermassive BHs are directly observed to be heating the hot gas in groups and clusters and the efficiency we have adopted has been constrained to yield a good match to the present-day black hole mass---halo mass relation and the cosmic black hole density.  In addition, as Dav{\'e} et al.\ point out, momentum-driven winds are incapable of shutting down cooling flows in the most massive haloes at the present-day, and therefore yield CBG star formation rates well in excess of what is observed.  Furthermore, this model does not appear to reproduce the large scatter seen in the X-ray luminosity-temperature relation of groups.
For these reasons, we argue that feedback from supermassive BHs is likely a key ingredient in galaxy group formation and evolution.  This certainly does not preclude, however, feedback from BHs and some form of momentum-driven winds working in tandem to shape the properties of groups.

Additional insights from observations of galaxy groups would be very helpful for discriminating between contending models and to further constrain the parameters of the subgrid physics.  In particular, extending entropy measurements down to lower system masses, where the differences between the models should be largest (e.g., Fig.\ 2), would be very useful.  Going beyond globally-averaged properties and azimuthally-averaged profiles and making detailed comparisons with the full 2D maps of groups (e.g., Finoguenov et al.\ 2006) also represents a very promising avenue for distinguishing between AGN heating mechanisms (e.g., by looking at the structure of cavities and sound-wave ripples).

In addition, adopting well-defined group selection functions for observed samples that can be applied directly to simulated catalogs would be helpful.  At present, most samples of groups with high quality X-ray observations do not have a well-defined selection function.  They are comprised of systems that have high enough surface brightnesses to extract radial profiles, but it is presently not clear whether these are `typical' systems.  In the case of massive galaxy clusters, for example, the central entropy varies by more than an order of magnitude at fixed halo mass (Cavagnolo et al.\ 2009).  The systems with the highest central entropies (longest cooling times) are referred to as non-cool core systems (or non-cooling flow systems, in the old jargon), and may represent more than 50\% of the cluster population.  The analogues of non-cool core clusters on the group scale have not yet been discovered, maybe because they do not exist, or perhaps, more interestingly, because their X-ray surface brightnesses are so low that they have escaped detection.  If a population of `X-ray dark groups' could be convincingly demonstrated to exist (e.g., in optically-selected group samples with follow-up X-ray observations), this would present a very interesting challenge to current cosmological simulations.

\section*{Acknowledgments}

The authors thank the anonymous referee for helpful suggestions that improved the paper.  The authors would also like to thank Ria Johnson, Ming Sun, and Jesper Rasmussen for providing 
their observational data, and Michael Balogh for helpful suggestions.  We thank Russell Smith and Patricia Sanchez-Blazquez for 
useful discussions regarding stellar metallicities.  IGM acknowledges support from a 
Kavli Institute Fellowship at the University of Cambridge.  RAC is supported by the Australian Research Council via a 
Discovery Project grant.  The simulations presented here were run on Stella, the LOFAR
BlueGene/L system in Groningen, on the Cosmology Machine at the
Institute for Computational Cosmology in Durham as part of the Virgo
Consortium research programme, and on Darwin in Cambridge. This work
was sponsored by National Computing Facilities Foundation (NCF) for
the use of supercomputer facilities, with financial support from the
Netherlands Organization for Scientific Research (NWO). 
This work was supported by Marie Curie Excellence Grant
MEXT-CT-2004-014112 and by an NWO VIDI grant.

\section*{Appendix: Effects of numerical resolution}

\begin{figure*}
\includegraphics[width=\columnwidth]{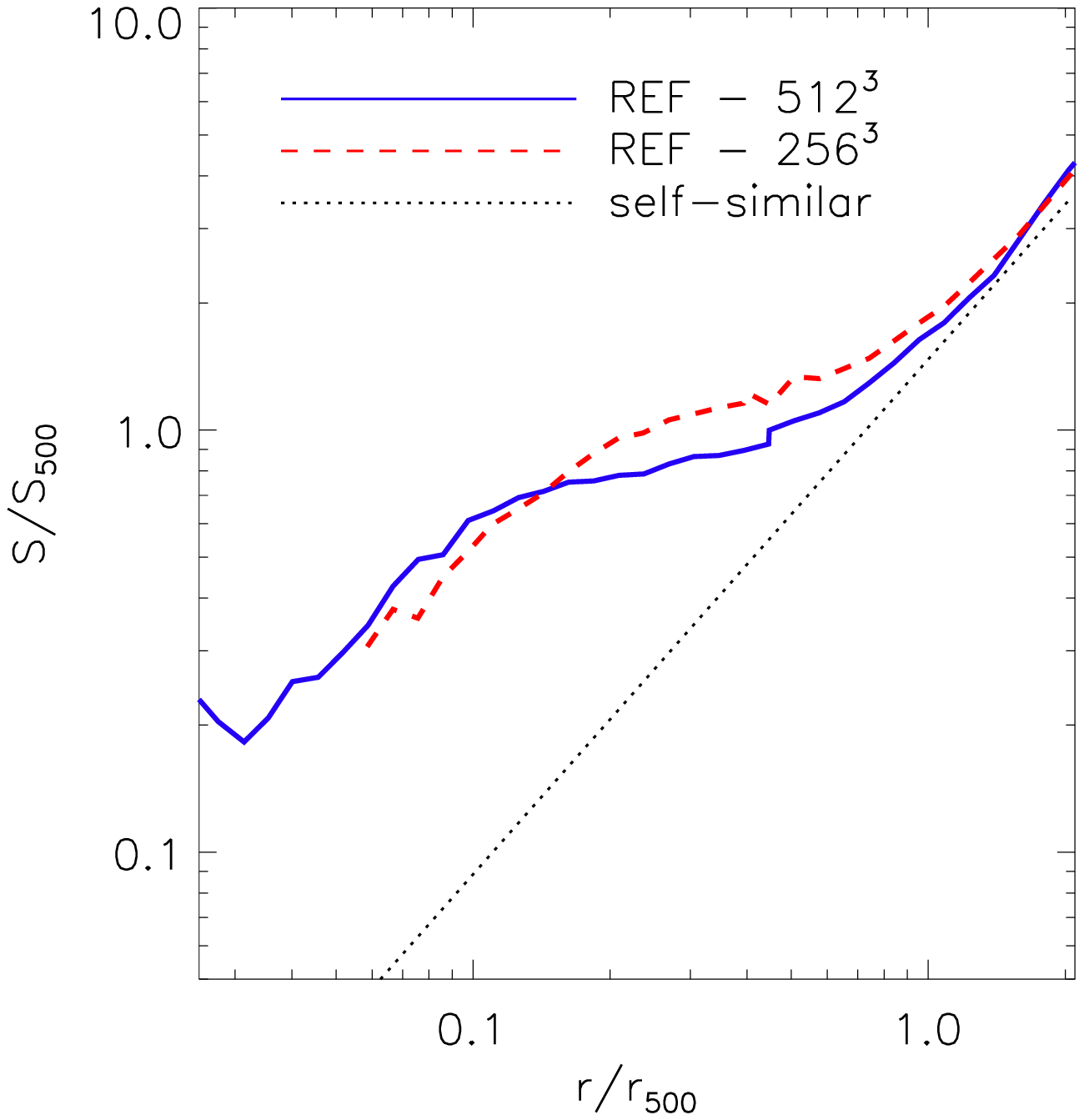}
\includegraphics[width=\columnwidth]{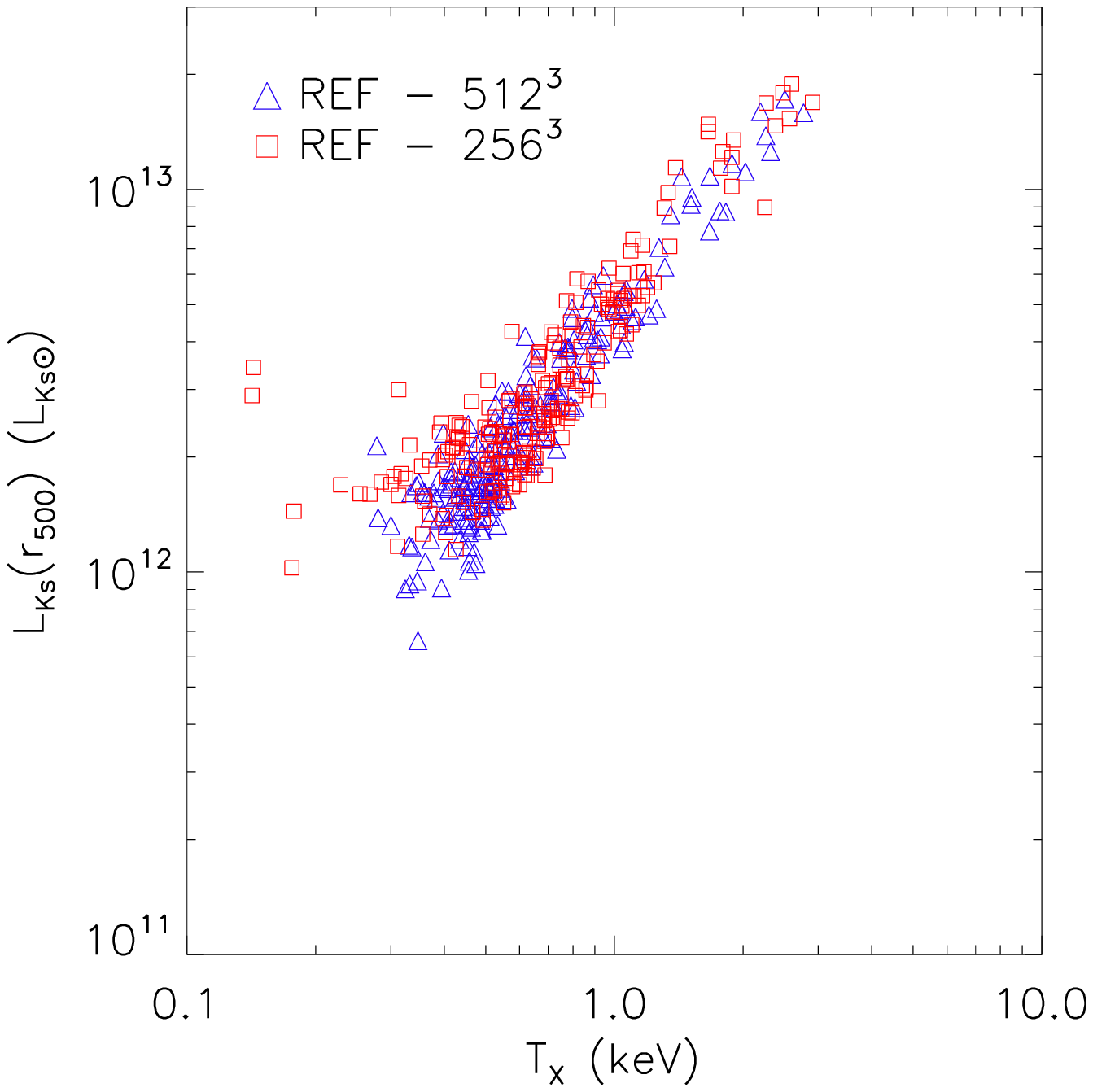}
\caption{The effect of numerical resolution on the median entropy profile and K-band luminosity - uncorrected mean temperature relation for the {\it REF} run.  (See Fig.\ 1 and Fig.\ 7 for comparison.)  Overall, the agreement between the two runs is excellent.}
\label{fig:res_test}
\end{figure*}

Here we explore the sensitivity of our results to numerical resolution.  The full production runs analysed in the main body of the paper contain $2\times512^3$ particles, yielding an (initial) baryon particle mass of $8.65 \times 10^7 h^{-1}$ M$_\odot$ and a dark matter particle mass of $4.06 \times 10^8 h^{-1}$ M$_\odot$ (note that due to stellar evolution the mass of gas/star particles can change during the course of a simulation, but typically only by a small amount).  We have also run simulations at lower resolution using $2\times256^3$ particles, implying the particle masses are increased by a factor of 8 over that in the full production run.

In Fig.\ 13 we compare the median entropy profiles for clusters with masses in the range $13.25 \le \log_{10}($M$_{500}/$M$_\odot) \le 14.25$ and the K-band luminosity - uncorrected mean temperature relations for the full production and lower resolution {\it REF} runs.  The results are robust to a decrease in the mass resolution (and we note that the small differences in the entropy profiles is well within the system-to-system scatter in the entropy profiles).  This indicates that the resolution of the production runs is adequate to robustly predict the properties of galaxy groups.

\bsp

\label{lastpage}

\end{document}